\documentclass[
    aps,
    prb,
    reprint,
    superscriptaddress,
    nofootinbib,
    amsmath,
    amssymb
]{revtex4-2}

\usepackage[english]{babel}
\usepackage[version=4]{mhchem}

\usepackage{amsmath,amssymb,bm,booktabs,mathtools}
\usepackage{graphicx}
\usepackage[compat=1.1.0]{tikz-feynman}
\tikzfeynmanset{warn luatex=false}
\usetikzlibrary{arrows.meta,decorations.markings,decorations.pathmorphing}
\tikzset{
    fermion/.style={
        line width=0.9pt,
        postaction={decorate},
        decoration={
            markings,
            mark=at position 0.56 with
                {\arrow{Stealth[length=2.2mm,width=1.5mm]}}
        }
    },
    cluster/.style={
        line width=1.4pt
    },
    interaction/.style={
        densely dashed,
        line width=1.0pt
    },
    vertex/.style={
        circle,
        fill=black,
        inner sep=1.15pt
    },
    lab/.style={
        font=\small
    },
    oplab/.style={
        font=\small,
        fill=white,
        inner sep=1.2pt
    }
}
\usepackage{bbold}
\usepackage[colorlinks=true, allcolors=blue]{hyperref}

\begin{document}

\title{Reduced scaling implementation of the coupled cluster-based static embedding method, MPCC with density fitting}

\author{Avijit Shee}\email{ashee@berkeley.edu}
\affiliation{Department of Chemistry, University of California, Berkeley, CA 94720, USA}
\author{Fabian M. Faulstich}\email{faulstich@rpi.edu}
\affiliation{Department of Mathematical Sciences, Rensselaer Polytechnic Institute, Troy, NY 12180, USA}
\affiliation{Department of Chemistry and Chemical Biology, Rensselaer Polytechnic Institute, Troy, NY 12180, USA}
\author{K. Birgitta Whaley}
\affiliation{Department of Chemistry, University of California, Berkeley, CA 94720, USA}
\affiliation{Berkeley Center for Quantum Information and Computation, Berkeley, CA 94720, USA}
\author{Lin Lin}
\affiliation{Department of Mathematics, University of California, Berkeley, CA 94720, USA}
\affiliation{Applied Mathematics and Computational Research Division, Lawrence Berkeley National Laboratory, Berkeley, CA 94720, USA}
\author{Martin Head-Gordon}
\affiliation{Department of Chemistry, University of California, Berkeley, CA 94720, USA}
\date{\today}

\begin{abstract}
An improved algorithm to implement our previously proposed static quantum embedding method, MPCC (J. Chem. Phys. 161, 164107 (2024)), is presented. It makes use of the density fitting (DF) approximation of the electron repulsion integrals (ERIs) for all layers of the embedding method, namely the low-level method, the screened interaction, and the high-level method. Additionally, an approximate factorized solution of the Sylvester equation is employed based on the Laplace transformation for the low-level method. To build the screened interaction we have explored an approximated construction of it based on the distinguishable cluster approximation (DCA) by Kats and Manby (J. Chem. Phys. 141, 061101 (2014)), which amounts to neglecting certain higher-order exchange and ladder diagrams. Combining all these techniques, we have obtained an $\mathcal{O}(N^4)$ algorithm for the low-level method; a non-iterative $ \mathcal{O}(N_{\rm frag} N^4)$ algorithm for the screened interaction; and an iterative $\mathcal{O}(N_{\rm frag}^6)$ algorithm for the fragment solver, where $N_{\rm frag}$ is the number of fragment orbitals and $N$ is the total number of orbitals. The resulting low-scaling implementation of the MPCC method is applied to trans polyacetylenes, that is, C$_{2n}$H$_{2n+2}$ molecules for chain lengths $n=1-8$, and the timing and accuracy of the method are compared to the original unfactorized implementation. Moreover, the approximations are further validated by applying the method to the potential energy surface of the \ce{O3} molecule. Finally, we apply the method to the selected molecules in the S22 dataset for which the performance of the MP2 method is known to be poor. The results demonstrate that the DF-based implementation of the MPCC method provides a significant reduction in computational cost while maintaining accuracy comparable to that of the original implementation.
\end{abstract}


\maketitle

\section{Introduction}

A central principle underlying reduced-scaling electronic structure methods is the nearsightedness of electron correlation, first articulated by Kohn \cite{kohn1996density}. Pulay's pioneering analysis of the localizability of dynamic electron correlation \cite{VRG:pulay:1983:CPL} provided an early foundation for exploiting this principle directly in correlated wave function calculations. In conventional local-correlation methods, the occupied orbitals are localized in real space, and compact virtual-orbital domains are assigned to individual occupied orbitals or orbital pairs. For sufficiently large, nonmetallic systems, the size of each local virtual space becomes asymptotically independent of the total system size, making linear or near-linear scaling algorithms possible. This field has since matured through major developments by Werner and co-workers \cite{werner2011efficient}, Neese and co-workers \cite{VRG:neese:2009:CP,VRG:neese:2011:JCTC}, Head-Gordon and co-workers \cite{Maslen1998Noniterative,Lee2000LocalTriatomics,Wang2023Sparsity}, and many others \cite{Nagy2016IntegralDirect,Kjaergaard2017DivideExpandConsolidate,Glasbrenner2020ReducedScaling,Li2023ClusterInMolecule}, and local-correlation techniques are now available in several widely used quantum-chemistry packages.

Quantum embedding methods \cite{Georges96,Kotliar06,dmet_knizia13,Zgid15,QDET_JCTC2022} exploit a complementary form of locality. Rather than constructing a virtual space for each occupied pair, they identify a localized fragment containing both occupied and virtual orbitals (typically low-energy) and treat that fragment with a high-level theory. The remainder of the system---the environment---is described at a lower level, while its influence on the fragment is retained through an effective bath, a screened interaction, a self-consistency condition, or some combination of these ingredients. The physical principle assumed here is that the most demanding short- and intermediate-range correlation effects can be resolved within a compact fragment, whereas the more spatially extended response of the environment can be captured economically by the low-level theory. Consequently, the dimension of the high-level problem can remain independent of the total system size. Partitioning the system into multiple fragments can reduce the size of each high-level calculation still further. The linear scaling algorithms are an uncoupled case of this multi-fragment embedding approach, in which each fragment contains an occupied pair and its associated virtual space.

Tensor factorization provides a second, complementary route to reducing computational cost. In density fitting (DF) \cite{VRG:beebe:1977:IJQC,VRG:lowdin:1965:JMP,Lowdin:2009:IJQC,Folkestad:2019:JCP}, also known as the resolution-of-the-identity (RI) approximation, the sparse four-index ERI tensor is represented approximately as sum of products of three-index tensors, where an auxiliary basis set is introduced to facilitate the factorization. This low-rank representation reduces both storage requirements and the cost of the dominant tensor contractions. Higher-order factorizations can provide further reductions. For example, tensor hypercontraction (THC) \cite{Hu2017ISDF,Lee2019ISDFTHC,Qin2023ISDFTheory,Parrish:2019:JCP} introduces a set of real-space interpolation or quadrature points, allowing the ERIs to be expressed in terms of two-index factors.

In this work, we combine quantum embedding and tensor factorization to approach the accuracy of a high-level calculation at a substantially reduced cost. We focus on MPCC, a static quantum embedding method introduced in our recent work \cite{Shee2024}. Motivated by diagrammatic perturbation theory and, more broadly, by the philosophy of dynamical mean-field theory (DMFT) \cite{Georges96}, MPCC retains a small set of low-order terms from the M{\o}ller--Plesset perturbation series to describe the environment, while treating the fragment with a partially resummed perturbation theory, namely coupled-cluster (CC) theory. Restricting the CC calculation to the fragment substantially reduces the dimension of the high-level problem. Achieving an overall reduction in computational cost, however, also requires a low-scaling treatment of the environment and of its coupling to the fragment.

In our original work \cite{Shee2024}, we noted that applying DF to the first-order dressed perturbation theory (MP1) (in this context, MP1 denotes the first-order correlation correction to the wave function) used as the low-level method would yield an $\mathcal{O}(N^5)$ algorithm. Here, we reduce this scaling further to $\mathcal{O}(N^4)$. Two developments make this reduction possible. First, we express the solution of the MP1-type Sylvester equation in a factorized form using a Laplace-transform representation. Second, we recast the feedback of the fragment on the low-level theory in a compact form. At fixed fragment size, the additional cost associated with this feedback grows only linearly with the total system size, although its prefactor depends on the size of the fragment.

The coupling of the environment to the fragment is captured by a static screened interaction in the MPCC theory. It is obtained by similarity transforming the bare Hamiltonian with the low-level cluster operator. A similar construction was used previously in Hamiltonian downfolding \cite{kowalski2018properties,kowalski2023sub,huang2023leveraging}, but has not been systematically incorporated into a self-consistent embedding framework. A central computational challenge of building static screened interaction is that the similarity transformation generates higher-than-two-body operators that are absent from the bare Hamiltonian. In MPCC, the use of CC as a fragment solver \cite{Shee2024,shee2026addressing} alleviates much of this difficulty, but constructing the screened interaction nevertheless remained one of the most expensive steps in the original implementation. Here, we develop an approximate construction of it. Motivated by the asymptotic separation of a fragment and its environment by a distance $R$, we neglect selected higher-order exchange and ladder contributions that decay more rapidly with $R$. This approximation substantially reduces the cost of constructing the screened interaction while retaining the dominant coupling terms.


The remainder of this article is organized as follows. Section~\ref{sec:notations} introduces the orbital indices used in the manuscript and their dimensions.  Section~\ref{sec:theory} briefly reviews the MPCC method and describes the new theoretical ideas behind simplifying the MP1-type low-level method. Section~\ref{sec:algo} presents the detailed DF-based algorithms for the low-level theory, the screened interaction, and the high-level fragment solver. Section~\ref{sec:results} reports numerical tests of the accuracy and efficiency of the resulting implementation. Finally, Sec.~\ref{sec:conclusion} summarizes our conclusions and outlines directions for future work.

\section{Notations} \label{sec:notations}

We will use spatial orbitals throughout this work. The following notations are used to denote the orbital indices and their dimensions:

\begin{itemize}
    \item Orbitals in the occupied space ($O$): $i, j, k, l, \hdots$
    \item Orbitals in the virtual space ($V$): $a, b, c, d, \hdots$
    \item Orbitals in the fragment occupied space ($O_F$): $I, J, K, L, \hdots$
    \item Orbitals in the fragment virtual space ($V_F$): $A, B, C, D, \hdots$
    \item Total number of orbitals: $N$
    \item Total number of occupied orbitals: $n_o$
    \item Total number of virtual orbitals: $n_v$
    \item Total number of auxiliary basis functions: $N_{\rm aux}$
    \item Total number of Laplace quadrature points: $n_{\rm L}$
    \item Total number of fragment orbitals: $N_{\rm frag}$
    \item Total number of fragment occupied orbitals: $n_{I}$
    \item Total number of fragment virtual orbitals: $n_{A}$
    \item Total number of environment occupied orbitals: $n_{i}$
    \item Total number of environment virtual orbitals: $n_{a}$   
\end{itemize}

\section{Theory} \label{sec:theory}

In the MP-CC approach, we partition the total orbital space into a set of (active) fragment (F) orbitals and a set of (inactive) environment (E) orbitals. The fragment orbitals describe the chemically relevant region requiring higher accuracy, spanning only a few atoms, localized bonds, or a chemically relevant subspace of the full basis set.
The environment orbitals, on the other hand, will describe the remainder of the system.
A valid partition of a given orbital set $\{\phi_p\}_{i=p}^K$ is specified by disjoint index sets that assign each occupied and virtual orbital to either the fragment or the environment.
Assuming an $N$-electron system, the occupied space is,
\begin{equation}
O = O_{\rm E} \cup O_{\rm F},
\end{equation}
where $O=\{\phi_i:1\le i\le N\}$ and
\begin{equation}
O_{\rm E}=\{\phi_i:i\in\mathcal{I}_{\rm occ}^{\rm E}\}, \qquad
O_{\rm F}=\{\phi_i:i\in\mathcal{I}_{\rm occ}^{\rm F}\},
\end{equation}
with $\mathcal{I}_{\rm occ}^{\rm E}\cap\mathcal{I}_{\rm occ}^{\rm F}=\emptyset$ and
$\mathcal{I}_{\rm occ}^{\rm E}\cup\mathcal{I}_{\rm occ}^{\rm F}=[\![N]\!] = \{1,...,N\}$.
The virtual space is partitioned similarly:
\begin{equation}
V=V_{\rm E}\cup V_{\rm F}, 
\end{equation}
where $V = \{ \phi_i : N < i \leq K \}$ and
\begin{equation}
V_{\rm E}=\{\phi_i:i\in\mathcal{I}_{\rm vir}^{\rm E}\}, \qquad
V_{\rm F}=\{\phi_i:i\in\mathcal{I}_{\rm vir}^{\rm F}\}
\end{equation}
with $\mathcal{I}_{\rm vir}^{\rm E}\cap\mathcal{I}_{\rm vir}^{\rm F}=\emptyset$ and $\mathcal{I}_{\rm vir}^{\rm E}\cup\mathcal{I}_{\rm vir}^{\rm F}=[\![K]\!] \setminus [\![N]\!].$
We denote the number of fragment and environment orbitals by $N_{\rm F}=|\mathcal{I}_{\rm occ}^{\rm F}|+|\mathcal{I}_{\rm vir}^{\rm F}|$ and $N_{\rm E}=|\mathcal{I}_{\rm occ}^{\rm E}|+|\mathcal{I}_{\rm vir}^{\rm E}|$, respectively. We slightly abuse the notation by using $O$ and $V$ to denote both the sets of occupied and virtual orbitals, respectively, and their cardinalities. The intended meaning should be clear from the context.

We note that the choice of basis to construct the fragment space plays a vital role in obtaining the desired accuracy with this method, similar to the other embedding approaches. In our previous works \cite{Shee2024,shee2026addressing,pierce2026cpd}, we have utilized the active valence active space \cite{sayfutyarova2017automated} (AVAS) scheme to construct this space, as it provides an automatic approach. Within the AVAS scheme, we aim to isolate a set of orbitals of interest (such as d-orbitals of a transition metal) from the full set of orbitals of a specific problem. Typically, a minimal atomic orbital basis (MINAO) is chosen to represent the region of interest, as it closely aligns with the idea of chemically relevant bonding and antibonding orbitals in the valence region. On the other hand, for the full calculation, we can (and should) choose a larger atomic orbital basis set.
A special characteristic of the AVAS scheme is that higher angular momentum atomic orbitals can be trivially added to the active space by choosing a slightly larger basis set than the minimal one. This feature allows us to capture the strongest dynamical correlations within the active space and also provides a means to increase the active space size systematically.     

Once a valid partition is made, we employ an exponential wavefunction parametrization,
\begin{equation}
|\Psi\rangle = e^{T}|\Phi_0\rangle, \qquad 
T=\sum_{n\ge 1} T_n ,
\end{equation}
with
\begin{equation}
\begin{aligned}
T_n
&= \frac{1}{(n!)^2} \sum_{i_1\cdots i_n}\sum_{a_1\cdots a_n}
t_{i_1\cdots i_n}^{a_1\cdots a_n}\,
a_{a_1}^\dagger\cdots a_{a_n}^\dagger a_{i_n}\cdots a_{i_1}\\
&=\sum_{i_1\cdots i_n}\sum_{a_1\cdots a_n}
t_{i_1\cdots i_n}^{a_1\cdots a_n}\,
X_{i_1\cdots i_n}^{a_1\cdots a_n},
\end{aligned}
\end{equation}
where $X$ are particle-hole excitation operators \cite{Cizek1966}. We define the fragment contribution at excitation rank $n$ as the restriction to fragment indices, i.e.,
\begin{equation}
T_n^{\rm F}=
\frac{1}{(n!)^2} \sum_{\substack{i_1,\ldots,i_n\in \mathcal I_{\rm occ}^{\rm F}\\
a_1,\ldots,a_n\in \mathcal I_{\rm vir}^{\rm F}}}
t_{i_1\cdots i_n}^{a_1\cdots a_n}\,
X_{i_1\cdots i_n}^{a_1\cdots a_n},
\end{equation}
and the environment contribution as its complement, i.e.,
\begin{equation}
T_n^{\rm E}=T_n-T_n^{\rm F}.
\end{equation}
Equivalently, $T^{\rm F}=\sum_{n\ge 1}T_n^{\rm F}$ and $T^{\rm E}=T-T^{\rm F}$. We denote by ${\bf t}^{\rm F}$ and ${\bf t}^{\rm E}$ the collections of amplitudes appearing in $T^{\rm F}$ and $T^{\rm E}$, respectively.
Note that $T^{\rm F}$ collects excitation operators that act {\it exclusively} within the fragment orbital subspace. In contrast, $T^{\rm E}$ comprises all remaining amplitudes, i.e., (i) ``purely'' environmental excitations involving only environment orbitals and (ii) ``mixed'' fragment-environment excitations. The mixed terms capture correlations that couple fragment and environment degrees of freedom, e.g., excitations from fragment occupied orbitals into environment virtual orbitals, or simultaneous excitations involving occupied orbitals from both regions.

In the spirit of quantum embedding, we will compute the cluster operators $T^{\rm F}$ and $T^{\rm E}$ at different levels of accuracy. Specifically, we determine the fragment cluster operator $T^{\rm F}$ using the standard CC formalism (the high-level, HL, theory in quantum embedding parlance), while treating the environment cluster operator $T^{\rm E}$ at a perturbative level (the lower-level, LL, theory).
Put differently, the similarity-transformed Hamiltonian that enters the CC projection equations is treated differently in the fragment and environment theories. For the fragment, we use the full similarity-transformed Hamiltonian, i.e.,
\begin{equation}
\overline{H}({\bf t}^{\rm F}; {\bf t}^{\rm E})
= e^{-T^{\rm E}-T^{\rm F}} H e^{T^{\rm F}+T^{\rm E}},
\end{equation}
whereas for the environment, we employ a lower-order perturbative approximation, denoted $\widetilde{H}({\bf t}^{\rm E}; {\bf t}^{\rm F})$. The latter is theoretically more involved; a detailed discussion of the low-level Hamiltonian proposed and investigated here is provided in Sec. \ref{sec:low_level_theory}. More specifically, we will discuss a convenient factorization of that equation that leads to a reduced scaling implementation.
In overview, the MP-CC approach leads to two sets of coupled projective equations
\begin{align}
     \langle \Phi_\mu^{\rm F}| \overline{H}^{\rm F}({\bf t}^{\rm F}; {\bf t}^{\rm E})| \Phi_0 \rangle = {}& 0, \label{Eq:projF} \\
     \langle \Phi_\mu^{\rm E}| \widetilde{H}^{\rm E}({\bf t}^{\rm E}; {\bf t}^{\rm F}) | \Phi_0 \rangle = {}& 0, \label{Eq:projEB}
\end{align}
where $|\Phi_\mu^{\rm F}\rangle$ and $|\Phi_\mu^{\rm E}\rangle$ denote excited Slater determinants in the fragment and environment space, respectively.
These coupled residual equations can be obtained from the stationarity of the MP-CC Lagrangian,
\begin{equation}
\label{Eq:LagrangianGen}
\begin{aligned}
\mathcal{L}({\bf t},\boldsymbol{\lambda}) 
&= \langle \Phi_0 | \overline{H} | \Phi_0 \rangle + \langle \Phi_0 | \Lambda^{\rm F} \overline{H}^{\rm F}({\bf t}^{\rm F}; {\bf t}^{\rm E}) | \Phi_0 \rangle\\
&\quad + \langle \Phi_0 | \Lambda^{\rm E} \widetilde{H}^{\rm E}({\bf t}^{\rm E}; {\bf t}^{\rm F}) | \Phi_0 \rangle, 
\end{aligned}    
\end{equation}
where $\Lambda^{Y}=\sum_{\mu\in Y}\lambda_\mu^{Y}X_\mu^{Y}$ for $Y\in\{{\rm F},{\rm E}\}$, and the first term in the Lagrangian is the coupled cluster energy for the total system (fragment + environment); therefore the $\overline{H}$ remained unlabeled.

We emphasize that both $\overline{H}^{\rm F}$ and $\widetilde{H}^{\rm E}$ in Eqs.~\eqref{Eq:projF} and~\eqref{Eq:projEB} depend on the full set of cluster amplitudes, i.e., ${\bf t}_{\rm F}$ {\it and} ${\bf t}_{\rm E}$. However, when solving the fragment residual equations, ${\bf t}_{\rm E}$ is held fixed and ${\bf t}_{\rm F}$ is updated, whereas in solving the environment residual equations, ${\bf t}_{\rm F}$ is held fixed and ${\bf t}_{\rm E}$ is updated. Consequently, the two sets of equations are coupled and can advantageously be self-consistently solved using a double-loop procedure consisting of \textit{macro}- and \textit{micro}-iterations.
In the \textit{macro}-iterative step, we first solve the low-level amplitude equations to obtain the environment cluster amplitudes, $T^{\rm E}$. We then construct the screened interaction for the fragment problem,
\begin{equation}
\label{eq:WF}
    W^{\rm F} = e^{-T^{\rm E}} H e^{T^{\rm E}},
\end{equation}
by similarity-transforming the bare Hamiltonian with $T^{\rm E}$ and subsequently restricting all indices to the fragment subspace. This operation typically generates effective Hamiltonian matrix elements with higher particle-rank (many-body) contributions, i.e., although $H$ is a two-body operator, $W^{\rm F}$ generally contains effective three-body and higher-body terms.
In the \textit{micro}-iterative step, we solve the fragment amplitude equations using $W^{\rm F}$. The resulting fragment amplitudes, ${\bf t}_{\rm F}$, enter the low-level amplitude equations and thereby update the next \textit{macro}-iterative step. We also note that $W^ {\rm F}$ is typically denoted as a downfolded Hamiltonian, and Eq.~\eqref{eq:WF} provides a static approach to build this quantity. A downfolded Hamiltonian can encapsulate the dynamical correlation effects of the environment into the fragment.  As the number of degrees of freedom of the fragment problem is much smaller than the total system, we can treat the downfolded Hamiltonian with a higher-accuracy solver. A similar static approach was taken previously by Kowalski \textit{et al.} to construct both unitary \cite{kowalski2018properties} and non-unitary \cite{kowalski2023sub} downfolded Hamiltonians, and Evangelista \textit{et al.} \cite{huang2023leveraging} used a different framework, namely the driven similarity renormalization group (DSRG) approach, for this construction. However, unlike our work, these studies solved the downfolded Hamiltonian in isolation, that is, not self-consistently within an embedding framework.

\subsection{Factorization of the doubles equation in low-level method} \label{sec:low_level_theory}

For the low-level method, we have explored a first-order screened perturbation theory as elaborated in our previous work \cite{Shee2024}. We will refer to this method as MP1. In the following, a theoretical procedure that leads to a low-scaling implementation of the doubles equation of the MP1 method is presented. The singles equation of the MP1 method can be solved with a computational cost of $\mathcal{O}(N^4)$ by using the DF integrals, and therefore we will not discuss it further.

We classify the environment double-excitation amplitudes into the following classes: $t_{ij}^{ab}$, $t_{ij}^{Ab}$, $t_{Ij}^{ab}$, $t_{iJ}^{aB}$, $t_{IJ}^{ab}$, $t_{ij}^{AB}$, $t_{iJ}^{AB}$, and $t_{IJ}^{aB}$.
The low-level projection equations used to determine these amplitudes are given by
\begin{equation}
    \langle \chi_{ij}^{ab, E} | W_{ij}^{ab, E} + [\tilde{F}, T_{2}] |\Phi_0\rangle = 0, \label{Eq:T2_ampl_env}
\end{equation}
where, $\tilde{F}$, $W_{ij}^{ab, E}$ are the one and two electronic part of the $\widetilde{H}^{\rm E}$, respectively. 

In these equations, we keep the active subset of the $T_2$ amplitudes, denoted by $T_{2,F}$, fixed and solve for the remaining classes of amplitudes. The $T_{2,F}$ amplitudes are obtained from a high-level equation and are therefore denoted by $T_{2}^{F, \rm HL}$. To enable a low-scaling evaluation of the environment amplitudes, we rewrite Eq.~\eqref{Eq:T2_ampl_env} as
\begin{equation}
\begin{aligned}
 \langle \chi_{ij}^{ab, E} | W_{ij}^{ab, E} + [\tilde{F}, T_{2}^{\rm LL}] + 
 \mathcal{R}
 |\Phi_0\rangle = 0, 
\label{Eq:T2_ampl_env_DC}
\end{aligned}
\end{equation}
%
where $\mathcal{R} =  [\tilde{F}, T_{2}^{F, \rm HL}] - [\tilde{F}, T_{2}^{F, \rm LL}]$; we note that the contribution from the doubly counted $T_{2}^{F, \rm LL}$ amplitudes is explicitly subtracted.
To obtain an initial estimate of the $T_{2}^{\rm LL}$ amplitudes from Eq.~\eqref{Eq:T2_ampl_env_DC}, we set $\mathcal{R} = 0$, that is, we assume $T_{2}^{F,\rm HL} = T_{2}^{F, \rm LL}$. This approximation leads to
\begin{equation}
   \langle \chi_{ij}^{ab} | W_{ij}^{ab} + [\tilde{F}, \overline{T}_{2}^{\rm LL}] |\Phi_0\rangle = 0.\label{Eq:T2_ampl_env_noDC}
\end{equation}
The solution $\overline{T}_{2}^{\rm LL}$ to Eq.~\eqref{Eq:T2_ampl_env_noDC} can be obtained via a non-iterative procedure with a computational cost of $\mathcal{O}(N^4)$ (vide infra). The error introduced into the $T_{2}^{\rm LL}$ amplitudes by the approximation $\mathcal{R} = 0$ can be estimated by subtracting Eq.~\eqref{Eq:T2_ampl_env_noDC} from Eq.~\eqref{Eq:T2_ampl_env_DC}, which yields
\begin{align}
    \langle \chi_{ij}^{ab} | [\tilde{F}, \Delta T_{2}^{\rm LL}] + \mathcal{R}  |\Phi_0\rangle = 0, \label{Eq:T2_error}
\end{align}
where $\Delta T_{2}^{\rm LL} = T_{2}^{\rm LL} - \overline{T}_{2}^{\rm LL}$ denotes the error in the low-level amplitudes.
An analysis of Eq.~\eqref{Eq:T2_error} shows that the $\mathcal{R}$ contributes only to corrections of the $t_{iJ}^{AB}$ and $t_{IJ}^{aB}$ classes of amplitudes, when we want to eliminate the leading-order error. See Appendix~\ref{App:SolutionToSylvesterEquation} for more details. As a result, Eq.~\eqref{Eq:T2_error} can be solved iteratively with a computational cost of $\mathcal{O}(N_{\rm frag}^4 n_v)$ and a storage requirement of $\mathcal{O}(N_{\rm frag}^3 n_v)$.

At this point, another important feature of Eq.~\eqref{Eq:T2_ampl_env} can be elaborated. Even if this is a perturbative equation, it would not diverge due to the appearance of a small denominator. This is because the $T_{2}^{\rm F}$ amplitudes are the ones for which the denominator can be small when the system has a small HOMO-LUMO gap. These amplitudes are evaluated from a non-perturbative high-level equation, and they are kept fixed in the low-level equation. For the environment amplitudes, at least one of the indices will be from the environment, and therefore the denominator will be large enough to ensure convergence of the perturbative equation.

We now describe how to obtain an $\mathcal{O}(N^4)$ algorithm for solving Eq.~\eqref{Eq:T2_ampl_env_noDC}; a detailed derivation is given in Appendix~\ref{App:SolutionToSylvesterEquation}. We note that Eq.~\eqref{Eq:T2_ampl_env_noDC} is equivalent to

\begin{equation}
\label{eq:StartingPoint}
   P(ij, ab) \left(
   \sum_c F_{bc}t^{ac}_{ij} - \sum_m F_{mi} t^{ab}_{mj}\right) =
   - \sum_Q \tilde{J}^Q_{ai} \tilde{J}^Q_{bj} 
\end{equation}
where the pair-exchange operator is defined by
$$
P(ij,ab)\sum_c F_{bc}t^{ac}_{ij}
=
\sum_c F_{bc}t^{ac}_{ij}
+
\sum_c F_{ac}t^{cb}_{ij}
$$
and 
\begin{equation}
P(ij,ab)\sum_m F_{mi}t^{ab}_{mj}
=
\sum_m F_{mi}t^{ab}_{mj}
+
\sum_m F_{mj}t^{ab}_{im}.
\end{equation}
Introducing the particle--hole index $ai$, we define
\begin{equation}
T_{ai,bj}
=
t^{ab}_{ij},
\qquad
S_{ai,bj}
=
\sum_Q \tilde{J}^Q_{ai}\tilde{J}^Q_{bj}.
\end{equation}
and the particle--hole Fock-difference matrix
\begin{equation}
K_{ai,ck}
=
F_{ac}\delta_{ik}
-
\delta_{ac}F_{ki}.
\end{equation}
Eq.~\eqref{eq:StartingPoint} therefore becomes
\begin{equation}
\label{Eq:Lyapunov}
KT
+
TK^\top
=
-S,
\end{equation}
which is a Lyapunov equation. Assuming that the spectrum of $K$ lies in the open right half-plane, the solution to Eq.~\eqref{Eq:Lyapunov} is given by
\begin{equation}
\label{Eq:T_laplace}
T
=
-\int_0^\infty
e^{-\tau K}
S
e^{-\tau K^\top}
\,d\tau .
\end{equation}
The matrix $K$ has the Kronecker-sum form 
\begin{equation} 
K = F^{\rm v}\otimes I_{\rm o} - I_{\rm v}\otimes \left(F^{\rm o}\right)^\top , 
\end{equation} 
where $F^{\rm v}_{ac} = F_{ac}$, and $F^{\rm o}_{ki} = F_{ki}$.
Therefore,
\begin{equation}
e^{-\tau K}
=
e^{-\tau F^{\rm v}}
\otimes
e^{+\tau \left(F^{\rm o}\right)^\top}
=:
G(\tau)
\otimes 
\overline{G}(\tau)
\end{equation}
This yields
\begin{equation}
\begin{aligned}
t^{ab}_{ij}
&=
T_{ai,bj}
\\
&=
-\int_0^\infty d\tau
\sum_Q
\sum_{ck,dl}
G_{ac}(\tau)
\overline{G}_{ki}(\tau)
\tilde{J}^Q_{ck}
\tilde{J}^Q_{dl}
G_{bd}(\tau)
\overline{G}_{lj}(\tau).
\end{aligned}
\end{equation}
Introducing the transformed three-index quantity
\begin{equation}
\label{Eq:Jhat_def}
\widehat{J}^Q_{ai}(\tau)
=
\sum_{ck}
G_{ac}(\tau)
\tilde{J}^Q_{ck}
\overline{G}_{ki}(\tau),
\end{equation}
we obtain
\begin{equation}
\label{Eq:t_final}
t^{ab}_{ij}
=
-\int_0^\infty d\tau
\sum_Q
\widehat{J}^Q_{ai}(\tau)
\widehat{J}^Q_{bj}(\tau).
\end{equation}
At this point, the remaining step is to approximate the Laplace integral by a numerical quadrature. Replacing the integral over $\tau$ by a finite sum over Laplace points $s_\mu$ with corresponding weights $w_\mu$ yields
\begin{equation}
\begin{aligned}
t_{ij}^{ab}
&\approx
-\sum_{\mu=1}^{n_{\rm L}} w_\mu
\sum_Q
\widehat{J}_{ai}^Q(s_\mu)\widehat{J}_{bj}^Q(s_\mu)\\
&= 
-\sum_{Q'} \widehat{J}_{ai}^{Q'} \widehat{J}_{bj}^{Q'},
\end{aligned}
\end{equation}
where we defined $Q' = (Q,s_\mu)$, and the transformed intermediates are given by
\begin{equation}
\widehat{J}_{ai}^{Q'}
=
\sum_{ck}
\sqrt{w_\mu}
G_{ac}(s_\mu)\tilde{J}_{ck}^Q\bar{G}_{ki}(s_\mu).
\end{equation}
This protocol has an overall computational cost of
\begin{equation}
\mathcal{O}\!\left(n_L N_{\rm aux} (n_v^2 n_o + n_v n_o^2)\right),
\end{equation}
and a memory requirement of
\begin{equation}
\mathcal{O}(N_{\rm aux} n_v n_o).
\end{equation}

\section{DF-based algorithm of the MPCC method:} \label{sec:algo}

 We have used the density fitted (DF) integrals in each section of the MPCC method. Besides, for each section further algorithmic improvements have been made to reduce the computational cost. In this section, we will describe all these aspects in detail.

\subsection{The Density Fitting Approximation}
The density fitting (DF) approximation of the electron repulsion integral (ERI) tensor is a standard tensor decomposition used in electronic structure methods.
Elements of the ERI tensor can be expressed in a basis of single particle function functions $\{ \phi_p \}_{p=1}^K$ as
\begin{equation}
\label{eq:g}
    G^{pq}_{st} = \iint \frac{\phi^*_{p}(r_1) \phi^*_q(r_2) \phi_s(r_1) \phi_t(r_2)}{\Vert r_1 - r_2 \Vert_2} dr_1 dr_2.
\end{equation}
The DF approximation decomposes the order-$4$ ERI tensor into the following low-rank representation
\begin{equation}
\label{eq:SQG}
    G^{pq}_{st} \overset{\mathrm{DF}}{\approx} \sum_{Q=1}^{N_{\rm aux}} J^{Q}_{ps} J^{Q}_{qt},
\end{equation}
where $N_{\rm aux}$ labels an optimized, predetermined auxiliary basis set which grows linearly with system size, i.e., typically between $2$ to $3$ times the dimension $V$.
We denote the tensor $J$ as the DF ERI tensor.
It should be noted that the DF like three center ERI tensors can also be constructed using the Cholesky decomposition~\cite{VRG:beebe:1977:IJQC,VRG:lowdin:1965:JMP,Lowdin:2009:IJQC,Folkestad:2019:JCP} technique.

The DF approximation formally reduces the computational storage complexity of the ERI tensor from $\mathcal{O}(N^4)$ to $\mathcal{O}(N^3)$. However, the DF approximation requires that the two indices associated with a given particle are represented using the {\it same} three-center DF tensor. This structural restriction is well known to limit the ability of the DF approximation to reduce the computational complexity of many high-scaling electronic structure methods. 
That said, methods which introduce the canonical polyadic decomposition (CPD) of the three-center ERI tensor using an analytic decomposition,\cite{Pierce:2025:JCTC:MP2,VRG:pierce:2021:JCTC,Pierce:2022:ETD,Pierce:2025:JCTC:CPB,VRG:benedikt:2011:JCP,Benedikt:2013:JCP,Benedikt:2013:MP,Bohm:2016:JCP,Schmitz:2017:JCP,Khoromskaia:2015:PCCP,Madsen:2018:JCP,Peng2017}, the pseudospectral\cite{VRG:friesner:1985:CPL,Friesner:1986:JCP,Langlois:1990:JCP,VRG:ringnalda:1990:JCP,Friesner:1991:ARPC,Martinez:1995:JCP,VRG:martinez:1992:JCP,Martinez:1994:JCP,Ko:2008:JCP,Martinez:1993:JCP} (PS) method, and the tensor hypercontraction\cite{VRG:hohenstein:2012:JCP,VRG:hohenstein:2012:JCPa,VRG:parrish:2012:JCP,Hohenstein:2013:JCP,VRG:parrish:2014:JCP,Shenvi:2013:JCP,Schutski:2017:JCP,Parrish:2019:JCP,Lee:2019:JCTC,VRG:hummel:2017:JCP,VRG:song::JCP,Hohenstein:2019:JCP,Hohenstein:2021:JCP,Hohenstein:2022:JCP,Jiang:2022:JCTC,Zhao:2023:JCTC,Datar:2024:JCTC,Schmitz:2017:JCP, Khoromskaia:2015:PCCP} (THC) approaches have shown to successfully reduce the computational complexity of many high-scaling electronic structure methods.

\subsection{Low-level method}

For reference, we use the term {\it unfactorized MP1} to denote the original dense implementation of the low-level equations, in which the four-index \(T_2^{\mathrm{\rm LL}}\) amplitudes are formed and updated explicitly rather than represented in the Laplace-factorized form introduced here. The factorized low-level kernel starts from the current singles amplitudes $t_i^a$, the density-fitted three-index integrals, and the high-level active doubles amplitudes $T_{2}^{F, \rm HL}$. In the first stage, we build the $T_1$-dependent intermediates 
\begin{equation*}
\begin{aligned}
X^Q_{ai}=\sum_b J^Q_{ab}t_i^b,\quad
X^Q_{ij}=\sum_a J^Q_{ai}t_j^a,\quad
X^Q=2\sum_{ia}J^Q_{ai}t_i^a,
\end{aligned}
\end{equation*}
and use them to form the transformed three-index quantities 
\begin{equation*}
\begin{aligned}
\widetilde J^Q_{ij} = J^Q_{ij} + X^Q_{ij},\quad {\rm and}\quad 
\widetilde J^Q_{ai} = J^Q_{ai} + X^Q_{ai} - \sum_j \widetilde J^Q_{ij} t_j^a,
\end{aligned}
\end{equation*}
together with the dressed Fock blocks $\widetilde F_{oo}$, $\widetilde F_{ov}$, and $\widetilde F_{vv}$, where
\begin{equation}
\begin{aligned}
\widetilde F_{ij}
  &= F_{ij} + \sum_Q J^Q_{ij}X^Q
  - \sum_{Qk} J^Q_{ki}X^Q_{kj},\\
  \widetilde F_{ab}
  &= F_{ab} + \sum_Q J^Q_{ab}X^Q
  - \sum_{Qk} J^Q_{ka}X^Q_{bk},\\
  \widetilde F_{jb}
  &= F_{jb} + \sum_Q J^Q_{bj}X^Q
  - \sum_{Qi} X^Q_{ji}J^Q_{bi}.
\end{aligned}
\end{equation}
The construction of the $T_1$-dependent intermediates has leading costs $\mathcal{O}(N_{\rm aux}n_{\rm o}n_{\rm v}^2)$ and $\mathcal{O}(N_{\rm aux}n_{\rm o}^2n_{\rm v})$. The construction of the dressed Fock blocks has the same fourth-order scaling, with the leading terms $\mathcal{O}(N_{\rm aux}n_{\rm o}^3)$, $\mathcal{O}(N_{\rm aux}n_{\rm o}n_{\rm v}^2)$, and $\mathcal{O}(N_{\rm aux}n_{\rm o}^2n_{\rm v})$ for the occupied--occupied, virtual--virtual, and occupied--virtual blocks, respectively. Thus, for $N_{\rm aux}\sim N$, this part of the algorithm scales as $\mathcal{O}(N^4)$.

The particle--particle and hole--hole Fock blocks are then shifted by the $\widetilde F_{ov}T_1$ terms to obtain the effective one-particle operators entering the Lyapunov equation,
\begin{equation}
\widehat F_{bc}
=
\widetilde F_{bc}
-
\sum_i \widetilde F_{ic} t_i^b,
\qquad
\widehat F_{ki}
=
\widetilde F_{ki}
+
\sum_c \widetilde F_{kc} t_i^c .
\end{equation}
This step costs $\mathcal{O}(n_{\rm o}n_{\rm v}^2+n_{\rm o}^2n_{\rm v})$, and is therefore lower order than the density-fitted intermediate construction. These matrices define the one-particle propagators that appear in Eq.~\eqref{Eq:t_final}.

We emphasize that the low-level doubles tensor is not formed explicitly. Instead, the Laplace representation of Eq.~\eqref{Eq:t_final} is evaluated in factorized form. For each Laplace point $s_\mu$ with weight $w_\mu$, the implementation applies the one-particle matrix exponentials $\exp(-s_\mu \widehat F_{vv})$ and $\exp(+s_\mu \widehat F_{oo})$ to the transformed three-index tensor $\widetilde J^Q_{ai}$. These exponentials are efficiently evaluated using a Chebyshev polynomial expansion. The Chebyshev matrix sequences for the scaled $\widehat F_{vv}$ and $\widehat F_{oo}$ blocks are constructed once and reused for all Laplace points. Their cost is $\mathcal{O}(K_{\rm v}n_{\rm v}^3+K_{\rm o}n_{\rm o}^3)$, where $K_{\rm v}$ and $K_{\rm o}$ are the required Chebyshev orders. For each Laplace point, the application of the two propagators to all auxiliary slices costs $\mathcal{O}(N_{\rm aux}n_{\rm o}n_{\rm v}(n_{\rm o}+n_{\rm v}))$. Therefore, the full factor construction scales as
\begin{equation}
\mathcal{O}\!\left[
n_{\rm L}N_{\rm aux}n_{\rm o}n_{\rm v}(n_{\rm o}+n_{\rm v})
+
K_{\rm v}n_{\rm v}^3
+
K_{\rm o}n_{\rm o}^3
\right],
\end{equation}
or $\mathcal{O}(N^4)$ for fixed $n_{\rm L}$ and fixed Chebyshev accuracy.

This gives the factor tensor
\begin{equation*}
Y^{Q\mu}_{ai}
=
\sqrt{w_\mu}
\sum_{ck}
\left[\exp(-s_\mu \widehat F_{vv})\right]_{ac}
\widetilde J^Q_{ck}
\left[\exp(+s_\mu \widehat F_{oo})\right]_{ki}.
\end{equation*}
The low-level doubles amplitudes are therefore represented only implicitly as
\begin{equation}
\label{eq:factorizedT2}
\overline t_{ij}^{ab}
=
-\sum_{Q\mu} Y^{Q\mu}_{ai} Y^{Q\mu}_{bj},
\end{equation}
which is the factorized approximation to Eq.~\eqref{Eq:T_laplace}. The storage requirement is $\mathcal{O}(n_{\rm L}N_{\rm aux}n_{\rm o}n_{\rm v})$, rather than $\mathcal{O}(n_{\rm o}^2n_{\rm v}^2)$. All subsequent contractions use this representation directly; the global four-index tensor $\overline T_2^{\rm LL}$ is not reconstructed.

The singles residual is evaluated by contracting the factorized doubles representation with $\widetilde F_{ov}$. The two doubles contributions that would contain $t_{ij}^{ab}$ in the dense implementation are rewritten as contractions over the compound factor index $(Q,\mu)$,
\begin{equation}
\begin{aligned}
\Omega_{ai}^{(2)}
&=
-2\sum_{Q\mu}
\left(
\sum_{bj} Y^{Q\mu}_{bj}\widetilde F_{jb}
\right)
Y^{Q\mu}_{ai}\\
&\quad+
\sum_{Q\mu j}
\left(
\sum_b Y^{Q\mu}_{bi}\widetilde F_{jb}
\right)
Y^{Q\mu}_{aj}.
\end{aligned}
\end{equation}
The first term requires only $\mathcal{O}(n_{\rm L}N_{\rm aux}n_{\rm o}n_{\rm v})$ operations, because the inner contraction produces one scalar for each compound factor index. The second term scales as $\mathcal{O}(n_{\rm L}N_{\rm aux}n_{\rm o}^2n_{\rm v})$. The remaining terms in the singles residual are evaluated from the transformed Fock and three-index intermediates in the usual way, and the total residual evaluation therefore remains fourth order.

If an active high-level doubles block is present, the kernel applies the correction implied by Eq.~\eqref{Eq:T2_error}. First, the low-level factorized approximation is restricted to the active--active block and compared with the high-level active amplitudes,
\begin{equation}
\label{eq:DeltaT_2F}
\Delta T_{2}^F
=
T_{2}^{F, \rm HL}
-
\overline T_{2}^{F, \rm LL}.
\end{equation}
Only this small active block is reconstructed, at a cost $\mathcal{O}(n_{\rm L}N_{\rm aux}n_I^2n_A^2)$ and storage $\mathcal{O}(n_I^2n_A^2)$, where $n_I$ and $n_A$ denote the numbers of active occupied and active virtual orbitals.

The correction equation is then projected onto the two boundary classes that couple to this mismatch, $t_{iJ}^{AB}$ and $t_{IJ}^{aB}$. All other classes are excluded explicitly, because their source terms vanish in Eq.~\eqref{Eq:T2_error}. The source construction scales as
\begin{equation}
\mathcal{O}(n_i n_I^2 n_A^2)
+
\mathcal{O}(n_a n_I^2 n_A^2),
\end{equation}
where $n_i$ and $n_a$ denote inactive occupied and inactive virtual dimensions. The retained correction tensors require only 
\begin{equation}
\mathcal{O}(n_i n_I n_A^2+n_a n_I^2 n_A)
\end{equation}
storage, i.e. $\mathcal{O}(N_{\rm frag}^3N)$.

The projected linear system is applied matrix-free: given trial tensors $\Delta t_{iJ}^{AB}$ and $\Delta t_{IJ}^{aB}$, the code evaluates the corresponding action of $[\widetilde F,\Delta T_2]$ within these two classes only, adds the source generated by $\Delta T_{2,F}$, and scales the residual by the usual orbital-energy denominator preconditioner. If the inactive-inactive Fock action is diagonal or semicanonical, each GMRES matrix-vector product scales as $\mathcal{O}(N_{\rm frag}^4N)$, and the projected solve scales as $\mathcal{O}(n_{\rm it}N_{\rm frag}^4N)$. If the full inactive-inactive Fock blocks are applied explicitly, the matvec additionally contains the dense inactive-space contractions $\mathcal{O}(n_i^2n_In_A^2+n_a^2n_I^2n_A)$. In either case, the solve is restricted to the two retained boundary classes and never acts on a global four-index doubles tensor. The resulting system is solved by GMRES, with the previous correction used as a warm start.

After convergence of the projected active correction, the tensors
$\Delta t_{iJ}^{AB}$ and $\Delta t_{IJ}^{aB}$ are not inserted into a global $T_2$ array. Instead, their contribution to the singles residual is added directly through the antisymmetrized contractions
\begin{equation}
\label{eq:updateOmega}
\begin{aligned}
\Omega_{Ai}
&=\Omega_{Ai} + 
\sum_{JB}
\left(
2\Delta t_{iJ}^{AB}
-
\Delta t_{iJ}^{BA}
\right)
\widetilde F_{JB},\\
\Omega_{aI}
&=\Omega_{aI} + 
\sum_{JB}
\left(
2\Delta t_{IJ}^{aB}
-
\Delta t_{JI}^{aB}
\right)
\widetilde F_{JB}.
\end{aligned}
\end{equation}
These contractions scale as $\mathcal{O}(n_i n_I n_A^2+n_a n_I^2 n_A)$, or $\mathcal{O}(N_{\rm frag}^3N)$. The residual of active singles is then projected out, and the remaining singles amplitudes are updated with the diagonal energy denominator at cost $\mathcal{O}(n_{\rm o}n_{\rm v})$. Thus the production kernel stores the doubles information as the Laplace factor tensor $Y$ plus the two projected active-boundary correction tensors, rather than as a dense four-index $T_2$ object. The dense reconstruction of $T_2$ is reserved for diagnostics and is not part of the low-scaling factorized algorithm. A summary of the contractions and their computational requirements can be found in Table~\ref{tabl:LLworking equations} in appendix~\ref{APP:pseudoCode}.
In the following, we use the term {\it factorized MP1} to denote this low-level approximation: the baseline amplitudes \(\bar T_2^{\mathrm{\rm LL}}\) are represented by the Laplace-factorized form of Eq.~\eqref{eq:factorizedT2}, while the difference between the high- and low-level fragment doubles is incorporated through the projected corrections in the \(t_{iJ}^{AB}\) and \(t_{IJ}^{aB}\) amplitude classes described in Eqs.~\eqref{eq:DeltaT_2F}--\eqref{eq:updateOmega}; no global four-index \(T_2^{\mathrm{\rm LL}}\) tensor is constructed.

\subsection{Algorithm for the screened interaction} \label{Sec:screened_algo}

The screened interaction is defined as:
\begin{equation}
    W^F = e^{-T^E} H e^{T^E}
\end{equation}

To build these tensors, a hierarchical strategy is employed, in which we first carry out the HT$_2$ contractions, and then another set of T$_2$ contractions on these tensors. We choose the contraction path such that no tensors of rank higher than 4 are generated. The details of the cost reduction will be discussed subsequently.

Let us first discuss the $HT_2$ group of terms appearing in the screened interaction. The most expensive terms in this group are the ladder diagrams, because they cannot be easily factorized using the DF approximation. Specifically, the compute cost of particle-particle ladder (PPL) term is $\mathcal{O}(n_o^2 n_v^4)$ and that of the particle-hole ladder term is $\mathcal{O}(n_o^3 n_v^3)$. However, because the indices of the resulting tensor $W^F$ are localized on the fragment, the compute scaling of those terms reduces to $\mathcal{O}(n_I^2 n_A^2 n_v^2)$ and $\mathcal{O}(n_I^2 n_A^2 n_v n_o)$, respectively, which is a significant reduction in the compute cost.  

The most expensive set of terms in the $W^F$ construction appears from the $HT_2^2$ contractions. To analyze these terms further, we have shown the explicit goldstone diagrams that arise from those contractions in Tab.~\ref{tab:rccd-diagrams} and the compute cost of them in Tab.~\ref{tab:cost_screened}. For terms A and B, we can see in Tab.~\ref{tab:rpa_diag} that they are easily factorizable when DF integrals are used, and the compute cost reduces to $\mathcal{O}(n_A n_I n_o n_v N_{\rm aux})$.

\begin{table}[t]
\centering
\caption{Goldstone diagrams A and B are easily factorizable. The red wavy line indicates the DF factorization of the two-electron integrals, which facilitates $\mathcal{O}(N^5)$ evaluation of these diagrams.}
\setlength{\tabcolsep}{2pt}
\renewcommand{\arraystretch}{1.05}
\begin{tabular}{@{}cc@{}}
\textbf{A} & \textbf{B} \\
\begin{tikzpicture}[x=0.78cm,y=0.78cm]
  \coordinate (Lext) at (-2.15,0);
  \coordinate (Lint) at (-0.65,0);
  \coordinate (Rint) at ( 0.65,0);
  \coordinate (Rext) at ( 2.15,0);
  \coordinate (VL) at (-0.65,1.55);
  \coordinate (VR) at ( 0.65,1.55);

  \draw[cluster]
    (Lext) -- (Lint)
    node[midway,below=5pt,oplab] {$T_2$};
  \draw[cluster]
    (Rint) -- (Rext)
    node[midway,below=5pt,oplab] {$T_2$};
  \draw[interaction]
    (VL) -- (VR)
    node[pos=0.25,above=5pt,oplab] {$\tilde v$};

  \draw[fermion]
    (Lext)
    .. controls (-2.55,0.45) and (-2.55,1.10)
    .. (-2.45,1.55);
  \draw[fermion]
    (-1.85,1.55)
    .. controls (-1.75,1.10) and (-1.75,0.45)
    .. (Lext);
  \node[lab,above] at (-2.45,1.55) {$a$};
  \node[lab,above] at (-1.85,1.55) {$i$};

  \draw[fermion]
    (Lint)
    .. controls (-1.05,0.45) and (-1.05,1.10)
    .. (VL);
  \draw[fermion]
    (VL)
    .. controls (-0.25,1.10) and (-0.25,0.45)
    .. (Lint);
  \node[lab,left=1pt] at (-1.03,0.88) {$c$};
  \node[lab] at (-0.50,0.58) {$k$};

  \draw[fermion]
    (Rint)
    .. controls (0.25,0.45) and (0.25,1.10)
    .. (VR);
  \draw[fermion]
    (VR)
    .. controls (1.05,1.10) and (1.05,0.45)
    .. (Rint);
  \node[lab] at (0.50,1.02) {$d$};
  \node[lab,right=1pt] at (1.03,0.88) {$l$};

  \draw[fermion]
    (Rext)
    .. controls (1.75,0.45) and (1.75,1.10)
    .. (1.85,1.55);
  \draw[fermion]
    (2.45,1.55)
    .. controls (2.55,1.10) and (2.55,0.45)
    .. (Rext);
  \node[lab,above] at (1.85,1.55) {$b$};
  \node[lab,above] at (2.45,1.55) {$j$};

  \foreach \p in {Lext,Lint,Rint,Rext,VL,VR}
    \node[vertex] at (\p) {};

  \draw[red,line width=1.1pt,decorate,
    decoration={snake,amplitude=1.2pt,segment length=4pt}]
    (0,0.95) -- (0,2.15);
\end{tikzpicture}
&
\begin{tikzpicture}[x=0.78cm,y=0.78cm]
  \coordinate (Lext) at (-2.15,0);
  \coordinate (Lint) at (-0.65,0);
  \coordinate (Rint) at ( 0.65,0);
  \coordinate (Rext) at ( 2.15,0);
  \coordinate (VL) at (-0.65,1.55);
  \coordinate (VR) at ( 0.65,1.55);

  \draw[cluster]
    (Lext) -- (Lint)
    node[midway,below=5pt,oplab] {$T_2$};
  \draw[cluster]
    (Rint) -- (Rext)
    node[midway,below=5pt,oplab] {$T_2$};
  \draw[interaction]
    (VL) -- (VR)
    node[pos=0.25,above=5pt,oplab] {$\tilde v$};

  \draw[fermion]
    (Lext)
    .. controls (-2.55,0.45) and (-2.55,1.10)
    .. (-2.45,1.55);
  \node[lab,above] at (-2.45,1.55) {$a$};

  \draw[fermion] (VL) -- (Lext);
  \draw[fermion]
    (-1.05,1.55)
    .. controls (-1.05,1.10) and (-1.05,0.45)
    .. (Lint);
  \draw[fermion]
    (Lint)
    .. controls (-0.25,0.45) and (-0.25,1.10)
    .. (VL);
  \node[lab,left=1pt] at (-1.30,0.88) {$k$};
  \node[lab,above] at (-1.05,1.55) {$i$};
  \node[lab] at (-0.50,0.58) {$c$};

  \draw[fermion]
    (Rint)
    .. controls (0.25,0.45) and (0.25,1.10)
    .. (VR);
  \draw[fermion] (VR) -- (Rext);
  \draw[fermion]
    (1.05,1.55)
    .. controls (1.05,1.10) and (1.05,0.45)
    .. (Rint);
  \node[lab] at (0.50,1.02) {$d$};
  \node[lab,right=1pt] at (1.50,0.88) {$l$};
  \node[lab,above] at (1.05,1.55) {$j$};

  \draw[fermion]
    (Rext)
    .. controls (2.55,0.45) and (2.55,1.10)
    .. (2.45,1.55);
  \node[lab,above] at (2.45,1.55) {$b$};

  \foreach \p in {Lext,Lint,Rint,Rext,VL,VR}
    \node[vertex] at (\p) {};

  \draw[red,line width=1.1pt,decorate,
    decoration={snake,amplitude=1.2pt,segment length=4pt}]
    (0,0.95) -- (0,2.15);
\end{tikzpicture}
\end{tabular}
\label{tab:rpa_diag}
\end{table}

However, for the diagrams (C-F), similar factorization is not possible, and we needed to adopt a different strategy, which we explain below. For diagram C, we have shown the contraction path to build the final HT$_2^2$ tensor that contributes to $W^F$ in Tab.~\ref{tab:dca-diagram-c}. In the intermediate HT$_2$ contraction step, the resulting tensor may have all the indices localized on the fragment, which we retain as a fragment tensor. Besides, it can generate tensors with maximally two indices on the environment, which again, with a final contraction with the second T$_2$ tensor, may generate a fragment tensor. However, the second type of intermediate tensor that contains the environment indices will have a compute cost of $\mathcal{O}(n_I n_A n_o^2 n_v^2)$ as opposed to the first type of intermediate tensor, which has a compute cost of $\mathcal{O}(n_I^2 n_A^2 n_o n_v)$. In this work, we neglect the second type of intermediate tensor and only retain the first type of intermediate tensor. Similar analysis can be performed for D, E and F diagrams, and we neglect the second type of intermediate tensor for those diagrams as well. The compute cost of the neglected intermediate tensors is shown in Tab.~\ref{tab:cost_screened}.

\begin{table}[t]
\centering
\caption{Goldstone diagram C. We have shown two contraction paths: HT$_2$ and HT$_2^2$. From the HT$_2$ path, we retain only that tensor for which all the indices are localized on the fragment. The red wavy line indicates where the second T$_2$ tensor is contracted.}
\setlength{\tabcolsep}{2pt}
\renewcommand{\arraystretch}{1.05}
\resizebox{\columnwidth}{!}{%
\begin{tabular}{@{}ccc@{}}

\begin{tikzpicture}[x=1.15cm,y=1.15cm]

  \coordinate (Lext) at (-2.15,0);
  \coordinate (Lint) at (-0.65,0);
  \coordinate (Rint) at ( 0.65,0);
  \coordinate (Rext) at ( 2.15,0);

  \coordinate (VL) at (-0.65,1.55);
  \coordinate (VR) at ( 0.65,1.55);

  \draw[cluster]
    (Lext) -- (Lint)
    node[midway,below=5pt,oplab] {$T_2$};

  \draw[cluster]
    (Rint) -- (Rext)
    node[midway,below=5pt,oplab] {$T_2$};

  \draw[interaction]
    (VL) -- (VR)
    node[midway,above=5pt,oplab] {$\tilde v$};

  \draw[fermion]
    (VL) -- (Lext);

  \draw[line width=0.9pt,
    postaction={decorate},
    decoration={
      markings,
      mark=at position 0.35 with
        {\arrow{Stealth[length=2.2mm,width=1.5mm]}}
    }]
    (Lint) -- (VR);

  \draw[line width=0.9pt,
    postaction={decorate},
    decoration={
      markings,
      mark=at position 0.72 with
        {\arrow{Stealth[length=2.2mm,width=1.5mm]}}
    }]
    (Rint) -- (VL);

  \draw[fermion]
    (VR) -- (Rext);

  \draw[red, line width=1.2pt, decorate,
    decoration={snake,amplitude=0.45mm,segment length=1.8mm}]
    (-0.22,0.775) -- (0.22,0.775);

  \draw[red, line width=1.2pt, decorate,
    decoration={snake,amplitude=0.45mm,segment length=1.8mm}]
    (1.18,0.775) -- (1.62,0.775);

  \draw[fermion]
    (Lext)
    .. controls (-2.55,0.45) and (-2.55,1.10)
    .. (-2.45,1.55);

  \draw[fermion]
    (-1.05,1.55)
    .. controls (-1.05,1.10) and (-1.05,0.45)
    .. (Lint);

  \draw[fermion]
    (1.05,1.55)
    .. controls (1.05,1.10) and (1.05,0.45)
    .. (Rint);

  \draw[fermion]
    (Rext)
    .. controls (2.55,0.45) and (2.55,1.10)
    .. (2.45,1.55);

  \node[lab,above] at (-2.45,1.55) {$a$};
  \node[lab,above] at (-1.05,1.55) {$i$};
  \node[lab,above] at (1.05,1.55) {$j$};
  \node[lab,above] at (2.45,1.55) {$b$};

  \node[lab,left=1pt] at (-1.30,0.88) {$k$};
  \node[lab]          at (-0.50,0.58) {$c$};
  \node[lab]          at (0.50,1.02) {$d$};
  \node[lab,right=1pt] at (1.50,0.88) {$l$};

  \foreach \p in {Lext,Lint,Rint,Rext,VL,VR}
    \node[vertex] at (\p) {};

\end{tikzpicture}
&
\begin{tikzpicture}[x=1.15cm,y=1.15cm]

  \coordinate (VL) at (-0.85,0);
  \coordinate (VR) at ( 0.85,0);

  \draw[line width=1.0pt,decorate,
    decoration={snake,amplitude=0.45mm,segment length=1.8mm}]
    (VL) -- (VR)
    node[midway,above=5pt,oplab] {$\tilde v_{\mathrm{eff}}$};

  \draw[fermion] (VL) -- (-1.35,-1.15);
  \draw[fermion] (-1.15,1.15) -- (VL);
  \node[lab,below] at (-1.35,-1.15) {$X= F, E$};
  \node[lab,above] at (-1.15,1.15) {$F$};

  \draw[fermion] (VR) -- (1.35,1.15);
  \draw[fermion] (1.35,-1.15) -- (VR);
  \node[lab,below] at (1.35,-1.15) {$X= F, E$};
  \node[lab,above] at (1.35,1.15) {$F$};

  \node[vertex] at (VL) {};
  \node[vertex] at (VR) {};

\end{tikzpicture}
&

\begin{tikzpicture}[x=1.15cm,y=1.15cm]

  \coordinate (VL) at (-0.85,0);
  \coordinate (VR) at ( 0.85,0);

  \draw[line width=1.0pt,decorate,
    decoration={snake,amplitude=0.45mm,segment length=1.8mm}]
    (VL) -- (VR)
    node[midway,above=5pt,oplab] {$\tilde v_{\mathrm{eff}}$};

  \draw[fermion] (VL) -- (-1.35,1.15);
  \draw[fermion] (-0.35,1.15) -- (VL);
  \node[lab,above] at (-1.35,1.15) {$F$};
  \node[lab,above] at (-0.35,1.15) {$F$};

  \draw[fermion] (VR) -- (0.35,1.15);
  \draw[fermion] (1.35,1.15) -- (VR);
  \node[lab,above] at (0.35,1.15) {$F$};
  \node[lab,above] at (1.35,1.15) {$F$};

  \node[vertex] at (VL) {};
  \node[vertex] at (VR) {};

\end{tikzpicture} \\

C & HT$_2$ & HT$_2^2$

\end{tabular}%
}
\label{tab:dca-diagram-c}
\end{table}

Interestingly, we have noticed that in the distinguishable cluster approximation (DCA) by Kats \textit{et.~al.}~\cite{katsDCA2013communication}, exactly the same diagrams, that is, (C-F) are neglected, but at the amplitude equation level. We will similarly denote this approximation as the DCA approximation as well. In the following, we will try to provide a physical picture of why these diagrams can be neglected for the construction of the screened interaction. In this analysis, the argument of fast multiple moment expansion (FMM) will be used to show that these diagrams are indeed negligible. We assume that the fragment is spatially localized, and the environment is $R$ distance away from the fragment and $R > r_{12}$, where $r_{12}$ is the interelectronic distance either on the fragment or the environment. For the foregoing discussion, it is necessary to analyze the distance dependence of two types of integral tensors that contribute to the diagrams (A-F) in Tab.~\ref{tab:rccd-diagrams}. The first type of integrals is the two-electron integrals with the following form:
\begin{equation}
\begin{aligned}
    \langle ij | ab \rangle 
    = \int \phi_i(r_1) \phi_a(r_1) \frac{1}{r_{12}} \phi_j(r_2) \phi_b(r_2)  dr_1 dr_2.
\end{aligned}
\end{equation}
Here, we can see that each vertex (ia or jb) can consist entirely of the fragment (F) orbitals or the environment (E) orbitals. Based on that we can classify the integrals as $V_{FF}$, $V_{EE}$, and $V_{FE}$. Among these classes of integrals, the $V_{FF}$ and $V_{EE}$ integrals will have a distance dependence of $\mathcal{O}(R^0)$. On the other hand, for the $V_{FE}$ integrals, the multipole expansion is valid, and the leading-order monopole-monopole contributions will vanish because of the orthogonality between the occupied $\phi_i$ and unoccupied $\phi_a$ orbitals. Therefore, the $V_{FE}$ integrals will have contributions, starting from the dipole-dipole interactions, and thus have a distance dependence of $\mathcal{O}(1/R^3)$. There will be another type of integrals for which a vertex can consist of one fragment orbital and one environment orbital. These integrals will decay as $\mathcal{O}(e^{-\gamma R})$. We will denote them as $V_{ex}$. 

The second type of integrals that need to be analyzed are the $T_2$ amplitudes, expressed as:
\begin{equation}
    T_{ij}^{ab} = \frac{\langle ij | ab \rangle}{\epsilon_i + \epsilon_j - \epsilon_a - \epsilon_b},
\end{equation}
where the denominator doesn't have an R-dependence. Therefore, the distance dependence of the $T_2$ amplitudes will be the same as that of the two-electron integrals.
The $T_2$ amplitudes therefore can be classified as $T_{FF}$, $T_{EE}$, and $T_{FE}$. The $T_{FF}$ and $T_{EE}$ amplitudes will have a distance dependence of $\mathcal{O}(R^0)$, while the $T_{FE}$ amplitudes will have a distance dependence of $\mathcal{O}(1/R^3)$. Similarly, we can also have $T_{ex}$ amplitudes that will decay as $\mathcal{O}(e^{-\gamma R})$, where $\gamma$ is a damping parameter.

While constructing the screened interaction, the $T_{FF}$ amplitudes will not contribute at all. Therefore, when analyzing diagram A, we can see that the leading-order contributions arise from the $T_{FE} V_{EE} T_{EF}$ term, and the distance dependence will be $\mathcal{O}(R^{-6})$.

Whereas, for the (B-E) diagrams there will be contributions from either $T_{ex}$ or $V_{ex}$ or from both; thus they will decay exponentially with distance, and finally for the diagram F, we will have the $T_{FE} V_{FE} T_{EF}$ term, which will decay as $\mathcal{O}(R^{-6})$. Now, we identify diagram A as one of the RPA terms, and because of its slow decay with distance, it will be the leading-order contribution to the screened interaction. Even though diagram B will decay exponentially, it is necessary to keep that diagram to maintain the fermionic exchange symmetry of the screened interaction \cite{katsDCAJCP2016}. The rest of the diagrams we neglect for their exponential decay with distance, except for diagram F. We neglect diagram F because the theory still remains exact for two-electron systems, even though it breaks the exchange symmetry of the screened interaction.

Here we also want to mention that the exclusion principle violating (EPV) diagrams also contribute to the $HT_2^2$ terms. However, the construction of those terms goes through a dressed Fock matrix like intermediate. Therefore, the computational costs of these diagrams are maximally $\mathcal{O}(N_{\rm aux} n_o^2 n_v^2)$ for a full coupled cluster calculation. However, in our scheme, these terms will contribute to: i) a fragment-only dressed Fock matrix, which will eventually feature in the fragment amplitude equations and ii) a fragment-environment type Fock matrix leading to a fragment-only residual contribution or to $W^F$. The computational cost of the type (ii) Fock matrix construction will be maximally $\mathcal{O}(N_{\rm aux} n_I n_o n_v^2)$. Although these terms are formally of type $\mathcal{O}(N^4)$, with a small prefactor of active occupied orbitals, they can be expensive due to the large number of auxiliary functions and inactive virtual orbitals. Therefore, we will report the timings for these diagrams separately in the results section. In passing, we also want to mention that the type (ii) Fock matrix we have mentioned above can also be neglected from the FMM argument we made earlier. But we have not explored that option in this work.  

Another analysis we want to put forward that justifies the neglect of the diagrams (C-F) is based on the energetic argument, hence more relevant when explicit spatial localization is not used to build the fragment. For term A, a close inspection would show that the T amplitudes that contribute to $W^F$ can have minimum one environment orbital. A similar analysis would show that the T amplitudes that contribute to the diagrams D, E, F will have at least two environment orbitals. Therefore, the energy denominator of the T amplitudes that contribute to the diagrams D, E, F will be larger than that of diagram A. Therefore, the contribution of diagrams D, E, F will be smaller than that of diagram A. 

Overall, with the DCA approximation, we can evaluate the screened interaction with a computational cost of $\sim$ $\mathcal{O}(N_{\rm {frag}}  N^4)$.

\begin{table}
    \centering
    \begin{tabular}{llc}
        \toprule
        \textbf{Step} & \textbf{Scaling} \\
        \midrule
        \quad H$T_2$: & \\
        1. Particle-Particle Ladder (PPL) \quad  & $n_{\text{v}}^2 n_{\text{A}}^2 n_{\text{I}}^2$ \\
        2. Particle-Hole Ladder (PHL) \quad  & $n_{\text{o}} n_{\text{v}}  n_{\text{A}}^2 n_{\text{I}}^2$ \\
        3. Hole-Hole Ladder (HHL) \quad  & $n_{\text{o}}^2 n_{\text{A}}^2 n_{\text{I}}^2$ \\
        \midrule
        \quad H$T_2^2$: & \\
        1. Particle-Hole ring (RPA) (A)  & $n_{\text{o}} n_{\text{v}} n_{\text{A}} n_{\text{I}}  n_{\text{aux}}$ \\
           and exchange (B) & \\
        2. Exclusion Principle Violating (EPV) & $ n_{\text{I}} n_{\text{o}} n_{\text{v}}^2 n_{\text{aux}}$ \\
        3. DCA terms (C-F) & $n_{\text{o}}^2 n_{\text{v}}^2 n_{\text{A}} n_{\text{I}}$ \\
        \bottomrule
    \end{tabular}
    \caption{Scaling of the most expensive terms in the screened interaction.}
    \label{tab:cost_screened}
\end{table}


\subsection{Algorithm for the high-level method} \label{Sec:high-level-algorithm}

The matrix elements of the screened fragment Hamiltonian, \(W_F\), are extracted using the algorithm described in the preceding section. The resulting Hamiltonian is represented by tensors of rank at most four whose indices are restricted entirely to the active fragment space. These tensors are subsequently used to solve the high-level problem at the CCSD level, for which the formal computational cost scales as \(\mathcal{O}(N_{\rm frag}^6)\).

Although the DCA approximation is employed in constructing the screened interaction, no analogous diagrammatic truncation is introduced in the high-level fragment calculation. Because the active fragment is spatially localized, its internal interactions are predominantly short-ranged, and the exchange and ladder contributions neglected in the DCA construction cannot be assumed to be small within the fragment. We therefore retain the complete set of CCSD diagrams in the high-level solver.

\section{Results and discussion} \label{sec:results}

We will numerically analyze both the performance and the accuracy of the low-scaling algorithm described in the previous sections. For the timing analysis, we will consider the trans polyacetylene (tPA) chain C$_{2n}$H$_{2n+2}$, with $n=1-8$, and for the accuracy analysis, we will consider the same tPA chain to assess the performance w.r.t. the increase of the system size.
It is important to note that, for the tPA chains considered here, the fragment size is not kept fixed as the total system size increases. Consequently, both $n_I$ and $n_A$ entering steps 7 and 8 of Appendix~\ref{APP:pseudoCode} increase. The reported scaling experiments therefore represent a particularly challenging test of MPCC, since the size of the physically relevant region grows together with the overall system rather than remaining constant. This additional fragment growth leads to a formally steeper asymptotic scaling than would be observed for a fixed-size fragment. By contrast, in less demanding scenarios, such as solvated systems, the fragment size may remain fixed while the environment grows through the inclusion of additional solvent shells. Nevertheless, our numerical results show that, over the range of systems investigated, the growth of the fragment is sufficiently modest that MPCC retains a favorable computational cost and enables fast simulations even for these increasingly demanding systems. We will furthermore consider the potential energy surface of the ozone molecule for the accuracy analysis. Finally some of the molecules from the S22 dataset \cite{jurecka2006benchmark_S22} will be studied for which the performance of the MP2 method is particularly challenging to assess the performance of the method for non-covalent interactions. 

\subsection{Timing analysis}
For the timing analysis, we choose the cc-pVTZ basis set for the full system and the STO-3G basis set for the AVAS based fragment construction. The actual size of the basis set and the fragment space have been provided in Tab.~\ref{tab:timing-orbital-sizes}. We have reported the timings for the low-level method and the screened interaction construction separately. For the screened interaction, the timings were obtained on a single core of an Intel(R) Xeon(R) Gold 6130 CPU @ 2.10GHz.

\begin{table}[t]
    \centering
    \begin{tabular}{rrrrr}
        \toprule
        system & $n_{\rm occ}$ & $n_{\rm vir}$& $n_{\rm active}^{\rm occ}$ & $n_{\rm active}^{\rm vir}$ \\
        \midrule
       \ce{C2H4} & 8  & 108 & 8 &  6    \\
       \ce{C4H6} & 15 & 189 & 15 & 11  \\
        \ce{C6H8} & 22 & 270 & 22 & 16  \\
        \ce{C8H10} & 29 & 351 & 29 & 21  \\
        \ce{C10H12} & 36 & 432 & 36 & 26  \\
        \ce{C12H14} & 43 & 513 & 43 & 31  \\
        \ce{C14H16} & 50 & 594 & 50 & 36  \\
        \ce{C16H18} & 57 & 675 & 57 & 41\\
        \bottomrule
    \end{tabular}
    \caption{Orbital and active-space sizes of the trans polyacetylene chains used in the timing analysis.}
    \label{tab:timing-orbital-sizes}
\end{table}

\subsubsection{Low-level method}

For each system, we have timed various sections of the low-level method. This includes one low-level update of the unfactorized MP1 reference and the factorized MP1 kernel. The factorized timings exclude any diagnostic reconstruction of the full $T_2$ tensor, so the reported cost reflects the production low-scaling workflow: the factorized $T_1$ residual construction, the Laplace factor build, and the projected active $T_2$ correction.

\begin{figure}[ht!]
    \centering
    \includegraphics[width=0.45\textwidth]{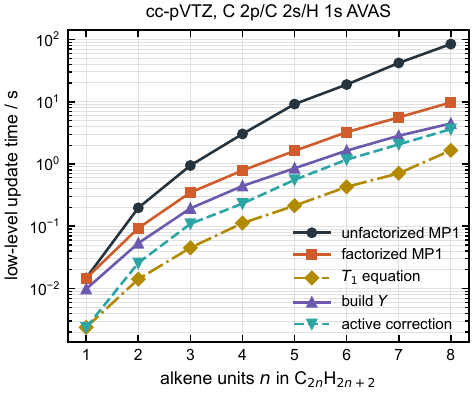}
    \caption{Low-level update timings for the unfactorized MP1 reference and the factorized MP1 kernel for the large valence AVAS active space. The factorized timing is decomposed into the $T_1$ equation, construction of the Laplace factor tensor $Y$, and the projected active $T_2$ correction.}
    \label{fig:ll_timing}
\end{figure}

\begin{figure}[ht!]
    \centering
    \includegraphics[width=0.4\textwidth]{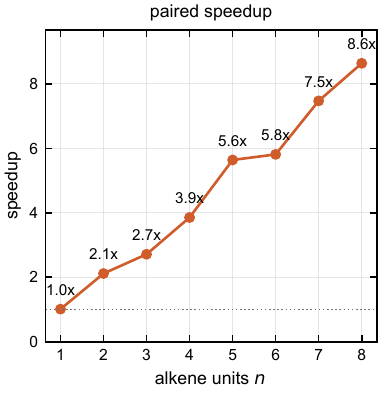}
    \caption{Paired speedup of the factorized MP1 low-level update relative to the unfactorized MP1 reference for the same alkene chains and active-space definition.}
    \label{fig:ll_timing_speedup}
\end{figure}

The timing decomposition in Fig.~\ref{fig:ll_timing} shows that the unfactorized MP1 update grows most rapidly with chain length, whereas the factorized kernel remains substantially cheaper. Within the factorized calculation, the construction of the Laplace factor tensor $Y$ is the dominant contribution for the larger chains, while the projected active correction remains below the full factorized update cost. The paired speedup in Fig.~\ref{fig:ll_timing_speedup} increases from essentially parity for ethene to about $5.8\times$ for C$_{12}$H$_{14}$, demonstrating that even for the large valence AVAS selection the factorized formulation recovers a clear timing advantage as the environment size grows.

\subsubsection{Screened interaction}
\begin{figure}
    \centering
    \includegraphics[width=0.5\textwidth]{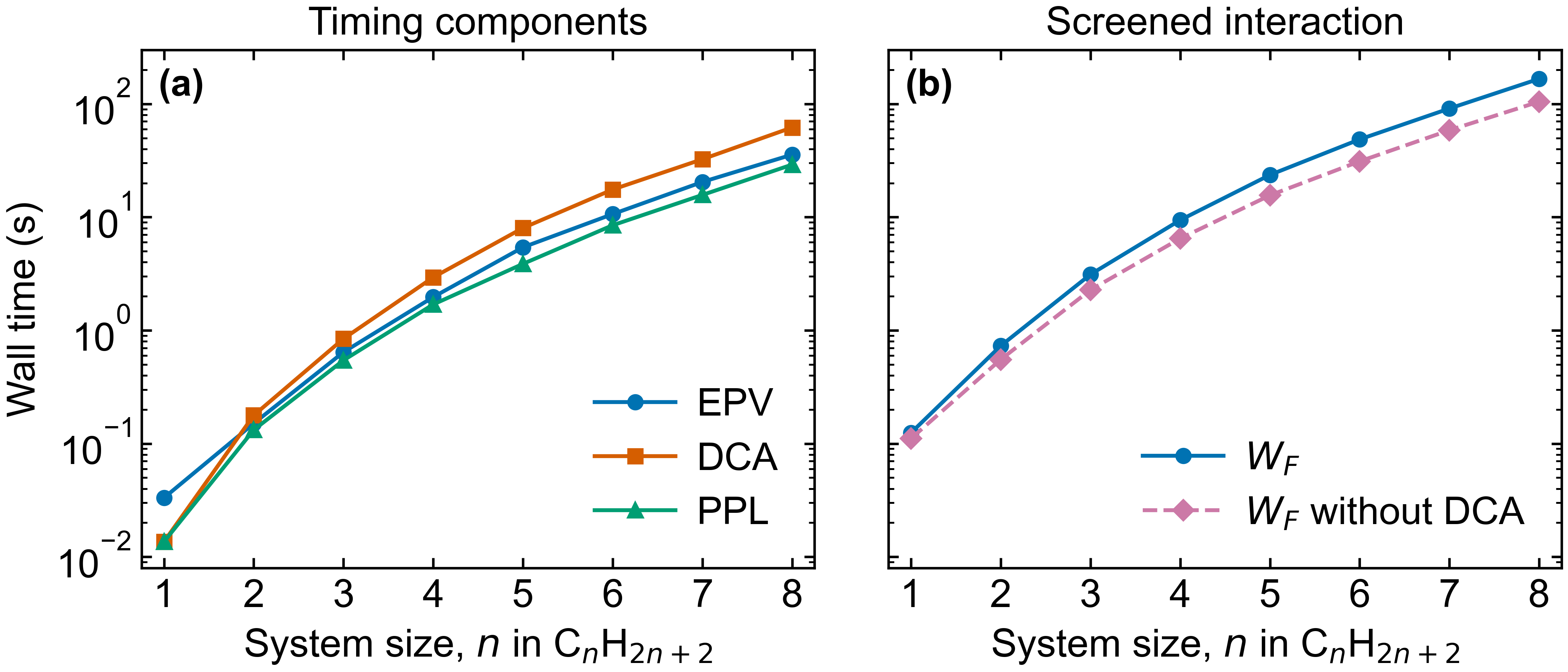}
    \caption{Timing for the screened interaction construction for the trans-polyacetylene oligomers. In panel (a), the timings of the most expensive terms, namely PPL, EPV, and DCA are shown. In panel (b), the total timing of constructing the screened interaction with and without the DCA approximation is shown.}
    \label{fig:screened_interaction}
\end{figure}

The timing for the screened interaction construction is shown in Fig.~\ref{fig:screened_interaction}. In panel (a) of this figure, we see the timing of the most expensive terms, namely PPL, EPV, and DCA, as we discussed before. According to the scaling analysis, the DCA terms dominate the construction of the screened interaction, and the cost of constructing the PPL terms is relatively small. From Tab.~\ref{tab:timing-orbital-sizes}, we can see that the increment in the number of total system orbitals (88) is much faster than the fragment orbitals (12) with the system size, therefore, the growth in timing of the PPL term is rather modest, which is otherwise the most expensive term in a CCSD calculation. The dependence of some of the screened interaction terms, such as PPL, particle-hole ladder, etc on the fragment size suggests that we should target to achieve a system size independent growth of the fragment space. In panel (b) we have shown the total timing of constructing the screened interaction with and without the DCA approximation. We can see that the DCA approximation is able to reduce the computational cost of constructing the screened interaction by a factor of $\sim 1.6$ for all the system sizes.

\subsection{Accuracy analysis}

\subsubsection{Peierls distortion of trans-polyacetylene:}

In quasi-one-dimensional conjugated polymers such as trans-polyacetylene (t-PA), the hypothetical metallic state is characterized by equidistant carbon-carbon bonds, which is electronically unstable. Through strong lattice distortions/electron-phonon coupling, the system undergoes a Peierls transition, a spontaneous symmetry breaking that leads to bond length alternation (BLA). The BLA is defined as the difference between adjacent long/single and short/double carbon-carbon bonds: $$\text{BLA} = R_{\rm C-C}^{\rm long} - R_{\rm C=C}^{\rm short}$$. This structural distortion opens a band gap at the Fermi level, lowering the total electronic energy and stabilizing the polymer into a semiconducting ground state. The energy difference between the metallic reference phase and the distorted ground state constitutes the dimerization energy, which is a fundamental metric for evaluating the strength of the Peierls instability.

We will plot the potential energy surface (PES) along the Peierls distortion coordinate or BLA for the C$_{\rm 2n}$H$_{\rm 2n+2}$ oligomers. For each chain length, geometries were optimized at the B3LYP/cc-pVDZ level by constraining the central bonds to discrete BLA values while allowing all remaining internal coordinates to relax, thereby minimizing chain-end geometric frustration. The undimerized, metallic reference geometry was obtained by imposing BLA = 0 throughout the conjugated backbone.

Single-point correlation calculations were then performed on these geometries using the cc-pVTZ basis set, with CCSD serving as the benchmark. MPCC calculations employed the same active space as in the timing analysis. To assess the effects of the different approximations on accuracy, we compared MPCC results obtained with and without the DCA approximation for the screened interaction and with factorized and non-factorized formulations of the low-level methods. MP2 calculations provided an additional reference for evaluating the improvements offered by MPCC.

For each constrained BLA point, the size-dependent dimerization energy per (CH)$_{\rm 2}$ monomer unit was evaluated as:
\begin{equation}
    \Delta \epsilon_{\rm dim} (n, \rm BLA) = \frac{E_{\rm BLA=0, n} - E_{\rm BLA, n}}{n},
\end{equation}

where E is the total electronic energy of the n-unit oligomer. The thermodynamic limit of the dimerization energy for each BLA was estimated by extrapolating the finite-size results to infinite chain length using a linear regression of $\Delta \epsilon_{\rm dim}$ versus $1/n$. 

\begin{equation}
    \Delta \epsilon_{\rm dim} (n, \rm BLA) = \Delta \epsilon_{\rm dim} (\infty, \rm BLA) + \frac{A}{n},
\end{equation}

The y-intercept of this regression, $\Delta \epsilon_{\rm dim} (\infty, \rm BLA)$, yields the dimerization energy per unit cell for the infinite polymer at that specific degree of structural distortion. Finally, to identify the global ground state of the bulk polymer, the extrapolated thermodynamic limit energies, $\Delta \epsilon_{\rm dim} (\infty, \rm BLA)$, were plotted as a function of BLA to locate the minimum energy configuration. The true macroscopic dimerization energy, alongside the optimal equilibrium BLA, was determined by locating the maximum of this extrapolated PES.

\begin{figure}
    \centering
    \includegraphics[width=0.5\textwidth]{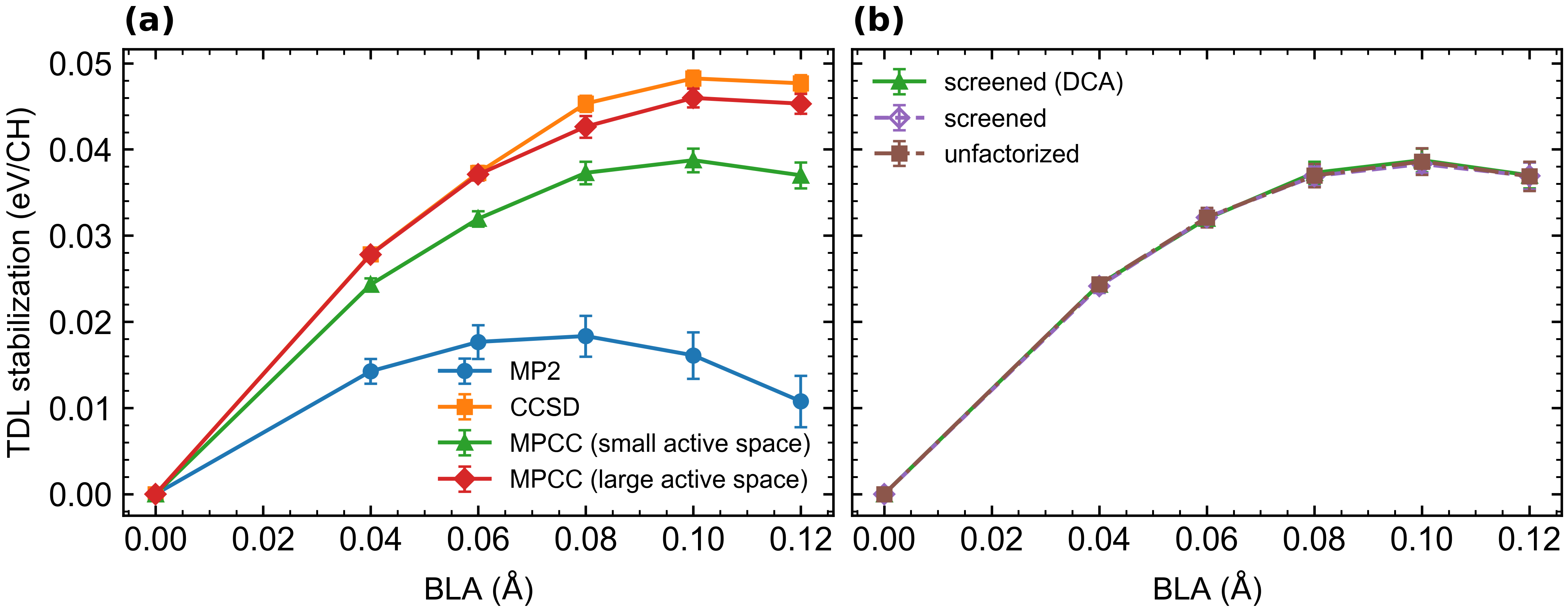}
    \caption{Dimerization energy per (CH)$_{\rm 2}$ unit as a function of bond length alternation (BLA) in the thermodynamic limit (TDL) of the trans-polyacetylene oligomer. In panel (a), the MPCC results are compared with the CCSD benchmark and MP2 results. In panel (a) small active space refers to the STO-3G basis set, and the large active space refers to the cc-pVDZ basis set. In panel (b), the effect of the DCA approximation on the screened interaction is assessed, and also the effect of the factorized low-level treatment is analyzed.}
    \label{fig:tPA_dimerization_accuracy}
\end{figure}

In panel (a) of Fig.~\ref{fig:tPA_dimerization_accuracy}, we have shown the dimerization energy per (CH)$_{\rm 2}$ unit as a function of BLA for the trans-polyacetylene oligomers with the CCSD, MP2, and MPCC methods. First, we observe that the MP2 method predicts a different optimal BLA value than CCSD (.08 $\AA$ instead of .10 $\AA$) and significantly underestimates the dimerization energy, predicting a value of 0.018 eV per (CH)$_{\rm 2}$ unit, which is only about 37.5\% of the CCSD reference value of .048 eV. In contrast, the MPCC method, even with a relatively small active space (STO-3G), recovers the same optimal BLA value as CCSD, and the dimerization energy it predicts is 80\% of the CCSD benchmark. The consideration of a larger active space (cc-pVDZ) in the MPCC calculation further improves the dimerization energy to 0.046 eV per (CH)$_{\rm 2}$ unit, which is 96\% of the CCSD reference value. This suggests that with the MP1 low-level treatment, the active space needs to be reasonably large for quantitatively accurate dimerization energy predictions.

In panel (b) of Fig.~\ref{fig:tPA_dimerization_accuracy}, we show that the inclusion of the DCA approximation in the screened interaction construction does not introduce any significant error, as evidenced by the near-identical results obtained with and without this approximation. Similarly, the factorized low-level treatment shows negligible deviation from the unfactorized approach, confirming that these approximations do not compromise accuracy for this system.

\subsubsection{Symmetric stretching of ozone molecule:}
\begin{figure}[h!]
    \centering
    \includegraphics[width=0.5\textwidth]{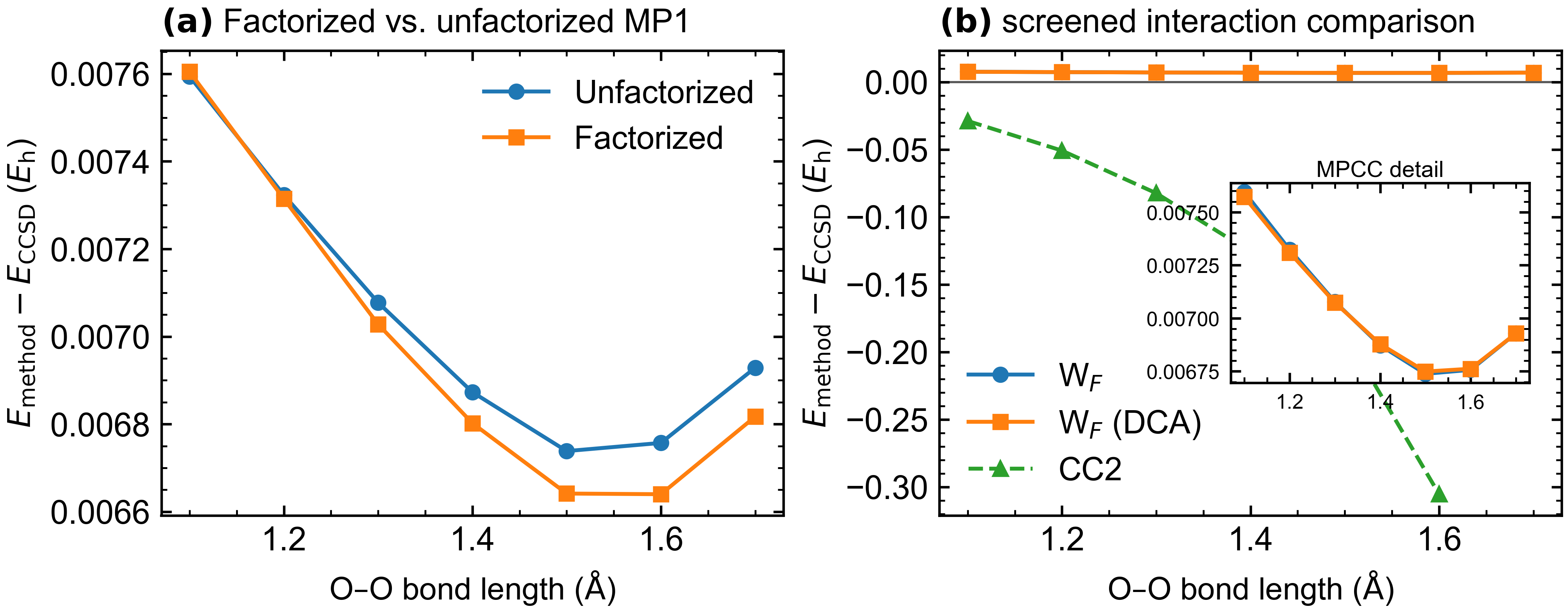}
    \caption{Various approximations used in the low-scaling implementation of the MPCC method have been compared with the potential energy surface of the ozone molecule. In panel (a), the effect of the factorized low-level treatment is assessed, and in panel (b), the effect of the DCA approximation on the screened interaction is analyzed.}
    \label{fig:ozone_pes}
\end{figure}

To assess the accuracy of the proposed approximations, we consider the symmetric stretching of the ozone molecule. All full-system calculations were performed using the aug-cc-pVTZ basis set, while the active fragment was constructed with the AVAS procedure using the cc-pVDZ basis as the minimal basis. Figure~\ref{fig:ozone_pes} shows the resulting potential-energy curves as functions of the common O--O bond length.

As previously observed and analyzed by Pabst \textit{et al.}~\cite{CC2_ozone:2010}, CC2 provides a qualitatively incorrect description of the ozone potential-energy curve. Upon stretching the O--O bonds, the HOMO--LUMO gap decreases, and the magnitudes of both the (T$_1$) and (T$_2$) amplitudes increase, signaling a near-degeneracy effect. Because CC2 treats the singles and doubles amplitudes at different perturbative orders, this regime shows an imbalance in their description and ultimately leads to the failure of the method.
However, with a small embedded space we can completely cure the failure of the CC2 method, while treating the large environment by the MP1 method as described in Sec.~\ref{sec:low_level_theory}.

Panel (a) of Fig.~\ref{fig:ozone_pes} assesses the effect of the factorized MP1 formulation by comparing it with the original, unfactorized MP1 treatment. The two formulations produce comparable errors relative to full-system CCSD over most of the potential-energy curve. At the largest O--O separations, the factorized formulation exhibits a modest increase in error, with a maximum additional deviation of approximately 0.2 $\mathrm{m}E_h$. This behavior signals that the contributions neglected in deriving the factorized approximation, as discussed in Sec.~\ref{sec:approx_low_level}, are reasonably small, but might be necessary for even higher accuracy.

Panel (b) examines the effect of the DCA approximation on the same symmetric-stretching coordinate. The deviations introduced by DCA remain negligible on the scale of the main figure throughout the range of bond lengths considered; the inset resolves that the residual error is (1e$^{-5}$ $E_h$), hence negligible. Thus, for the ozone potential-energy curve, the DCA approximation substantially simplifies the construction of the screened interaction without introducing an appreciable loss of accuracy.

\subsection{S22 Dataset:}
\begin{figure}
    \centering
    \includegraphics[width=0.5\textwidth]{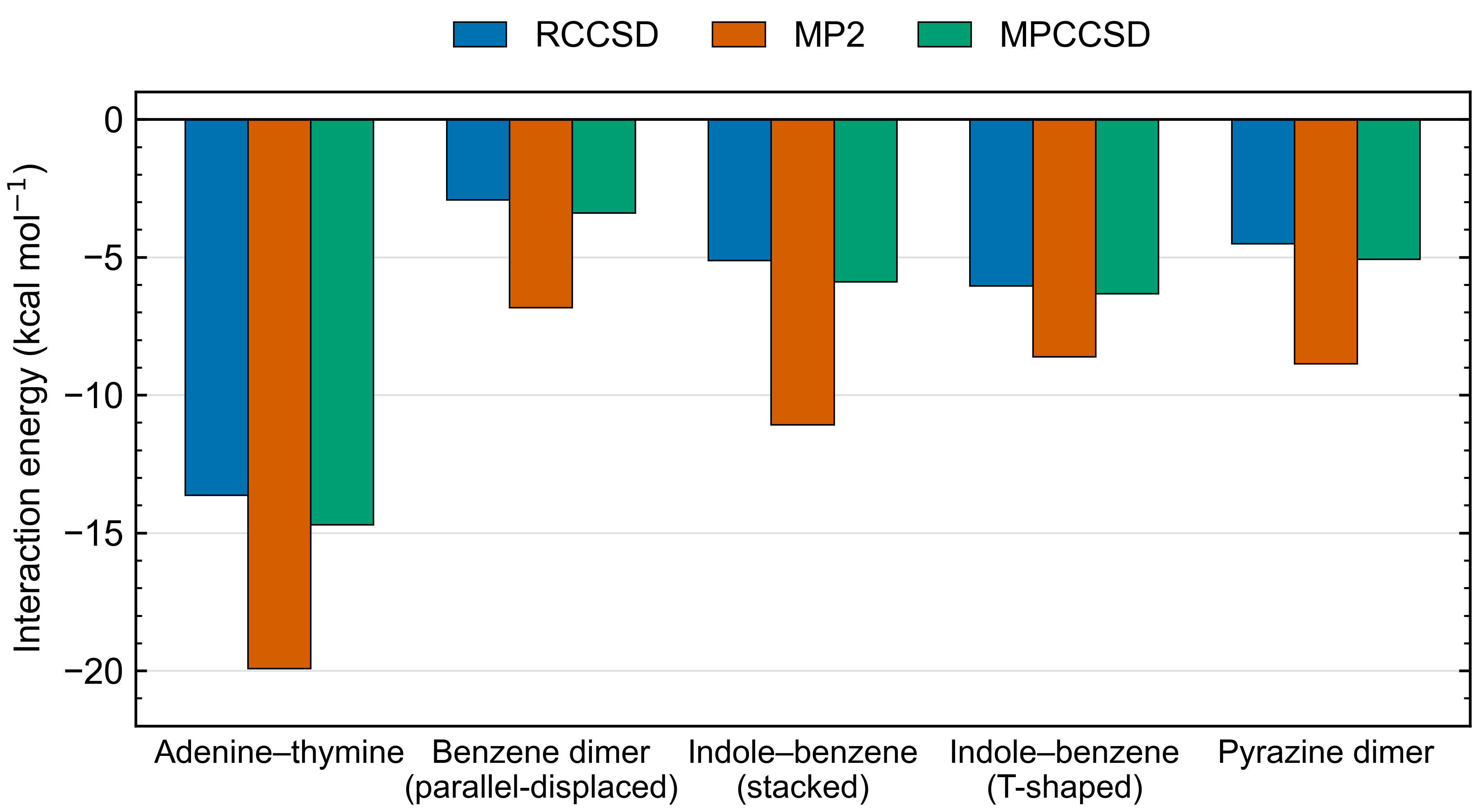}
    \caption{Interaction energies of the selected set of molecules in the S22 dataset were compared at the CCSD, MP2 and MPCC levels of theory. For the MPCC calculations, the def2-TZVPD basis set was used for the full system, and the def2-SVP basis set was used for the fragment. The DCA approximation was used for the construction of the screened interaction, and the factorized MP1 equation described in Sec.~\ref{sec:low_level_theory} was used to solve the low-level projection equations.}
    \label{fig:s22_results}    
\end{figure}

The S22 dataset \cite{jurecka2006benchmark_S22} is a benchmark set of 22 non-covalent complexes, which are commonly used to evaluate the performance of methods in describing weak interactions. This dataset consists of various molecular systems of polar, dispersive, and hydrogen-bonded nature. We will use this dataset to assess the accuracy of our method with only a handful of systems for which the MP2 method is known to perform poorly. With this consideration, we studied the following systems: (1) benzene-indole T-shape, (2) benzene-indole stack, (3) adenine-thymine complex stack, (4) benzene dimer T-shaped, (5) pyrazine dimer. Although the CCSD method is not considered a benchmark method for the S22 dataset, we will use it as a reference to assess the performance of our method. For the MPCC calculations, we used the def2-TZVPD basis set for the full system and the def2-SVP basis set for the fragment. For fragment space construction, we ignore the 1s core for each of the atoms in the second row of the periodic table. The DCA approximation was used for the construction of the screened interaction, and the factorized MP1 equation described in Sec.~\ref{sec:low_level_theory} was used to solve the low-level projection equations. The results of our calculations are summarized in Fig.~\ref{fig:s22_results}. The results show that the MPCC method is able to recover the CCSD interaction energies with a mean absolute error of 0.63 kcal/mol and a maximum absolute error of 1.07 kcal/mol. 

\section{Conclusions} \label{sec:conclusion}

In this work, we have developed a low-scaling algorithm for the MPCC method, so that the method can be used reliably for large systems. The low-scaling algorithm uses the 3-index density-fitted integrals instead of the 4-index integrals throughout and addresses the computational bottleneck of the three distinct steps of the MPCC method: (1) the low-level projection equations, (2) the construction of the screened interaction, and (3) the high-level method. The computational scaling of the low-level projection equations is reduced from $\mathcal{O}(N^5)$ to $\mathcal{O}(N^4)$ by solving the Sylvester equation in a factorized form using the Laplace transformation. The construction cost of the screened interaction has also been reduced to $ \mathcal{O}(N_{\rm frag} N^4)$ by neglecting the most expensive diagrams in the HT$_2^2$ contractions, which is justified by the fast decay of the transformed fragment-environment interactions. The approximation used for the screened interaction is similar to the distinguishable cluster approximation (DCA) by Kats \textit{et al.} \cite{katsDCA2013communication}. The high-level method is solved using the screened interaction, and its computational cost is $\mathcal{O}(N_{\rm frag}^6)$ and therefore ideally it is independent of the actual system size.

The theoretical reduction in computational cost has been verified by numerical experiments on the trans polyacetylene (tPA) chain C$_{2n}$H$_{2n+2}$, with $n=1-8$. The results show that the factorized low-scaling algorithm can achieve an $\sim \mathcal{O}(N)$ improvement over the unfactorized implementation of the low-level projection equations. The accuracy of the factorized low-scaling algorithm has been assessed by studying the Peierls stabilization energy of the tPA oligomers, potential energy surface of the ozone molecule and the interaction energies of the S22 dataset. The results show that the approximations are quite robust for the tPA chain, are qualitatively correct for the ozone molecule, and can recover the CCSD interaction energies with a mean absolute error of 0.63 kcal/mol for the S22 dataset. However, it is still necessary to include the higher-order error corrections in the low-level projection equations as suggested in Appendix~\ref{sec:approx_low_level} to achieve exact quantitative accuracy of the unfactorized implementation. The approximations made in the screened interaction on the other hand has been shown to incur negligible penalty in accuracy for the ozone molecule and the S22 dataset, and thus can be used reliably for large systems. The neglected diagrams in the HT$_2^2$ contractions account for 1/3 of the total computational cost of the screened interaction, and therefore their exclusion speeds up the calculation of the screened interaction by a factor of $\sim 1.6$ for the tPA chain. Also, we have shown that the most expensive particle-particle ladder (PPL) term in regular CCSD implementation is relatively inexpensive in our method.  

Finally, we want to remark that the speedup of the overall algorithm depends on the ratio $\mathcal{O}(N_{\rm frag}/N)$. For example, the speedup of the PPL term depends quartically on that ratio. Therefore, we will explore avenues to reduce the size of the fragment space in future work. For instance, explicit spatial localization of the fragment orbitals \cite{spade_mayhall_2019,senjean2021_IFO} can be used to reduce the size of the fragment space further on top of the overlap based AVAS scheme. Another approach to reduce the size of the fragment space is to use an improved low-level method than the MP1 method used here. The efficiency of the low-level method can be further improved by adopting recently developed linear-scaling strategies for the perturbative methods \cite{Wang2023Sparsity}. Also, the computational cost of the fragment solver scales linearly with the number of fragments, and therefore the overall speedup of the algorithm can be improved by using multiple fragments. We will explore these avenues in future work.

\section{Acknowledgments}
This material is based upon work supported by the U.S. Department of Energy, Office of Science, Office of Advanced Scientific Computing Research and Office of Basic Energy Sciences, Scientific Discovery through Advanced Computing (SciDAC) program under Award Number DE‐SC0022198 (A.S., K.B.W. and L.L.), and Award Number DE-SC0022364 (M.H-G.). 
F.M.F.~acknowledges support from the National Science Foundation under CAREER Grant No.~DMS-2543961. Any opinions, findings, conclusions, or recommendations expressed
in this material are those of the author and do not necessarily reflect the views of the National Science Foundation. Calculations were performed using the Savio computational cluster resource provided by the Berkeley Research Computing program at the University of California, Berkeley.

\appendix

\onecolumngrid

\section{Factorization of the low-level method:} \label{sec:approx_low_level}
In Sec.~\ref{sec:low_level_theory}, we have described how to obtain an $\mathcal{O}(N^4)$ algorithm for solving the low-level projection equations. In this section, we will point out an approximation made in this derivation. In Eq.~\eqref{Eq:T2_error}, we have shown that the error in the low-level amplitudes can be obtained by solving an equation that involves only the $t_{iJ}^{AB}$ and $t_{IJ}^{aB}$ classes of amplitudes. This is because the term in parentheses in Eq.~\eqref{Eq:T2_error} can generate terms that contribute to the correction of these classes of amplitudes, which can be seen from the active index restrictions in the $T_{2,F}$ amplitudes and the explicitly connected terms in the commutator $[\tilde{F}, T_{2,F}]$. Since there are no index restrictions on $\tilde{F}$, we can generate at max one inactive index in the correction term. Now the new $t_{iJ}^{AB}$ and $t_{IJ}^{aB}$ set of amplitudes can in turn generate corrections to the amplitudes that can maximally have two inactive indices and so on. But these corrections will be of second and higher order to the error; therefore, we can neglect them.

\section{Solution to Sylvester equation}
\label{App:SolutionToSylvesterEquation}
We start from
\begin{equation}
\label{Eq:R2_start}
P(ij,ab)\sum_c F_{bc}t^{ac}_{ij}
-
P(ij,ab)\sum_m F_{mi}t^{ab}_{mj}
+
\sum_Q \tilde{J}^Q_{ai}\tilde{J}^Q_{bj}
=
0 .
\end{equation}
The simultaneous pair-exchange operator is defined such that
\begin{equation}
\label{Eq:P_virtual}
P(ij,ab)\sum_c F_{bc}t^{ac}_{ij}
=
\sum_c F_{bc}t^{ac}_{ij}
+
\sum_c F_{ac}t^{cb}_{ij}
\end{equation}
and 
\begin{equation}
\label{Eq:P_occupied}
P(ij,ab)\sum_m F_{mi}t^{ab}_{mj}
=
\sum_m F_{mi}t^{ab}_{mj}
+
\sum_m F_{mj}t^{ba}_{mi}
=
\sum_m F_{mi}t^{ab}_{mj}
+
\sum_m F_{mj}t^{ab}_{im}.
\end{equation}
Therefore,
\begin{equation}
\label{Eq:R2_expanded}
\begin{aligned}
-\sum_Q \tilde{J}^Q_{ai}\tilde{J}^Q_{bj}
&=
\sum_c F_{bc}t^{ac}_{ij}
+
\sum_c F_{ac}t^{cb}_{ij}
-
\sum_m F_{mi}t^{ab}_{mj}
-
\sum_m F_{mj}t^{ab}_{im}.
\end{aligned}
\end{equation}
The natural compound index is the particle--hole index $ai$, not the
separate pair indices $ab$ and $ij$. We therefore define
\begin{equation}
\label{Eq:T_S_def}
T_{ai,bj}
=
t^{ab}_{ij},
\qquad
S_{ai,bj}
=
\sum_Q \tilde{J}^Q_{ai}\tilde{J}^Q_{bj}.
\end{equation}
The spatial-orbital pair symmetry yields
\begin{equation}
\label{Eq:T_symmetry}
T_{ai,bj}
=
t^{ab}_{ij}
=
t^{ba}_{ji}
=
T_{bj,ai}.
\end{equation}
Then we introduce the particle--hole Fock-difference matrix
\begin{equation}
\label{Eq:K_def}
K_{ai,ck}
=
F_{ac}\delta_{ik}
-
\delta_{ac}F_{ki}.
\end{equation}
Then
\begin{equation}
\label{Eq:KT_expand}
\begin{aligned}
(KT)_{ai,bj}
&=
\sum_{ck}
K_{ai,ck}T_{ck,bj}
\\
&=
\sum_{ck}
\left(
F_{ac}\delta_{ik}
-
\delta_{ac}F_{ki}
\right)
T_{ck,bj}
\\
&=
\sum_c F_{ac}t^{cb}_{ij}
-
\sum_k F_{ki}t^{ab}_{kj}.
\end{aligned}
\end{equation}
Furthermore,
\begin{equation}
\label{Eq:TKt_expand}
\begin{aligned}
(TK^\top)_{ai,bj}
&=
\sum_{ck}
T_{ai,ck}K_{bj,ck}
\\
&=
\sum_{ck}
T_{ai,ck}
\left(
F_{bc}\delta_{jk}
-
\delta_{bc}F_{kj}
\right)
\\
&=
\sum_c F_{bc}t^{ac}_{ij}
-
\sum_k F_{kj}t^{ab}_{ik}.
\end{aligned}
\end{equation}
Combining Eqs.~\eqref{Eq:KT_expand} and~\eqref{Eq:TKt_expand}, Eq.~\eqref{Eq:R2_expanded} becomes
\begin{equation}
KT
+
TK^\top
=
-S.
\end{equation}
Assuming that the spectrum of $K$ lies in the open right half-plane, the solution is
\begin{equation}
T
=
-\int_0^\infty
e^{-\tau K}
S
e^{-\tau K^\top}
\,d\tau .
\end{equation}
To verify this, note that
\begin{equation}
\label{Eq:derivative_identity}
\frac{d}{d\tau}
\left(
e^{-\tau K}
S
e^{-\tau K^\top}
\right)
=
-
K e^{-\tau K}S e^{-\tau K^\top}
-
e^{-\tau K}S e^{-\tau K^\top}K^\top .
\end{equation}
Hence,
\begin{equation}
\label{Eq:laplace_verification}
\begin{aligned}
KT+TK^\top
&=
-\int_0^\infty
\left(
K e^{-\tau K}S e^{-\tau K^\top}
+
e^{-\tau K}S e^{-\tau K^\top}K^\top
\right)
\,d\tau
\\
&=
\int_0^\infty
\frac{d}{d\tau}
\left(
e^{-\tau K}
S
e^{-\tau K^\top}
\right)
\,d\tau
\\
&=
\left[
e^{-\tau K}
S
e^{-\tau K^\top}
\right]_{\tau=0}^{\tau=\infty}
\\
&=
-S.
\end{aligned}
\end{equation}
To obtain the Kronecker representation of $K$, we regard $ai$ and $ck$
as composite particle--hole indices. With the convention
\begin{equation}
(A\otimes B)_{ai,ck}=A_{ac}B_{ik},
\end{equation}
the two terms in Eq.~\eqref{Eq:K_def} can be identified separately. First,
\begin{equation}
\label{Eq:K_first_kron}
F_{ac}\delta_{ik}
=
\left(F^{\rm v}\otimes I_{\rm o}\right)_{ai,ck}
\end{equation}
where
\begin{equation}
\label{Eq:Fv_def}
F^{\rm v}_{ac}=F_{ac}.
\end{equation}
Second,
\begin{equation}
\label{Eq:K_second_kron}
\delta_{ac}F_{ki}
=
\left(I_{\rm v}\otimes \left(F^{\rm o}\right)^\top\right)_{ai,ck}
\end{equation}
where the occupied Fock block is defined by
\begin{equation}
\label{Eq:Fo_def}
F^{\rm o}_{ik}=F_{ik}.
\end{equation}
Indeed,
\begin{equation}
\label{Eq:Fo_transpose_component}
\left[\left(F^{\rm o}\right)^\top\right]_{ik}
=
F^{\rm o}_{ki}
=
F_{ki}.
\end{equation}
Therefore, the element-wise definition
\begin{equation}
\label{Eq:K_component_repeat}
K_{ai,ck}
=
F_{ac}\delta_{ik}
-
\delta_{ac}F_{ki}
\end{equation}
is equivalent to the Kronecker-sum form
\begin{equation}
\label{Eq:K_kron}
K
=
F^{\rm v}\otimes I_{\rm o}
-
I_{\rm v}\otimes \left(F^{\rm o}\right)^\top .
\end{equation}

Now, using the property of the Kronecker product we can show that
\begin{equation}
\label{Eq:kron_exp}
\left [F^{\rm v}\otimes I_{\rm o}, I_{\rm v}\otimes \left(F^{\rm o}\right)^\top \right ]
=
0.
\end{equation}

Therefore,
\begin{equation}
\label{Eq:K_exp_factor}
e^{-\tau K}
=
e^{-\tau F^{\rm v}}
\otimes
e^{+\tau \left(F^{\rm o}\right)^\top}.
\end{equation}
Define
\begin{equation}
\label{Eq:G_def}
G_{ac}(\tau)
=
\left[
e^{-\tau F^{\rm v}}
\right]_{ac},
\qquad
\overline{G}_{ki}(\tau)
=
\left[
e^{+\tau F^{\rm o}}
\right]_{ki}.
\end{equation}
Then
\begin{equation}
\label{Eq:K_exp_indices}
\left[
e^{-\tau K}
\right]_{ai,ck}
=
G_{ac}(\tau)\overline{G}_{ki}(\tau).
\end{equation}
Using
\begin{equation}
\label{Eq:S_factorized}
S_{ck,dl}
=
\sum_Q
\tilde{J}^Q_{ck}
\tilde{J}^Q_{dl},
\end{equation}
Eq.~\eqref{Eq:T_laplace} gives
\begin{equation}
\label{Eq:t_laplace_expanded}
\begin{aligned}
t^{ab}_{ij}
&=
T_{ai,bj}
\\
&=
-\int_0^\infty d\tau
\sum_Q
\sum_{ck,dl}
G_{ac}(\tau)
\overline{G}_{ki}(\tau)
\tilde{J}^Q_{ck}
\tilde{J}^Q_{dl}
G_{bd}(\tau)
\overline{G}_{lj}(\tau).
\end{aligned}
\end{equation}
Introducing the transformed three-index quantity
\begin{equation}
\label{Eq:Jhat_def_appendix}
\widehat{J}^Q_{ai}(\tau)
=
\sum_{ck}
G_{ac}(\tau)
\tilde{J}^Q_{ck}
\overline{G}_{ki}(\tau),
\end{equation}
we obtain
\begin{equation}
t^{ab}_{ij}
=
-\int_0^\infty d\tau
\sum_Q
\widehat{J}^Q_{ai}(\tau)
\widehat{J}^Q_{bj}(\tau).
\end{equation}

\section{Pseudo Code}
\label{APP:pseudoCode}

\begin{table}[ht!]
  \centering
  \begin{tabular}{rll}
  \toprule
  \toprule
  \textbf{Step} & \textbf{Description} & \textbf{Scaling} \\
  \midrule
  \midrule

  1. & Compute $X^Q_{ai}$, $X^Q_{ij}$, and $X^Q$ & $\mathcal{O}(N^4)$ \\
  \hline
  & $X^Q_{ai} = \sum_b J^Q_{ab} t_i^b$ &
  $\mathcal{O}(n_{\rm o} n_{\rm v}^2 N_{\rm aux})$ \\
  & $X^Q_{ij} = \sum_a J^Q_{ai} t_j^a$ &
  $\mathcal{O}(n_{\rm o}^2 n_{\rm v} N_{\rm aux})$ \\
  & $X^Q = 2\sum_{ai} J^Q_{ai} t_i^a$ &
  $\mathcal{O}(n_{\rm o} n_{\rm v} N_{\rm aux})$ \\

  \vspace{-2mm}\\
  2. & Compute transformed DF intermediates $\widetilde J^Q$ &
  $\mathcal{O}(N^4)$ \\
  \hline
  & $\widetilde J^Q_{ij} = J^Q_{ij} + X^Q_{ij}$ &
  $\mathcal{O}(n_{\rm o}^2 N_{\rm aux})$ \\
  & $\widetilde J^Q_{ai}
  = J^Q_{ai} + X^Q_{ai}
  - \sum_j \widetilde J^Q_{ij} t_j^a$ &
  $\mathcal{O}(n_{\rm o}^2 n_{\rm v} N_{\rm aux})$ \\

  \vspace{-2mm}\\
  3. & Compute $\widetilde F$ blocks &
  $\mathcal{O}(N^4)$ \\
  \hline
  & $\widetilde F_{ij}
  = F_{ij} + \sum_Q J^Q_{ij}X^Q
  - \sum_{Qk} J^Q_{ki}X^Q_{kj}$ &
  $\mathcal{O}(n_{\rm o}^3 N_{\rm aux})$ \\
  & $\widetilde F_{ab}
  = F_{ab} + \sum_Q J^Q_{ab}X^Q
  - \sum_{Qk} J^Q_{ka}X^Q_{bk}$ &
  $\mathcal{O}(n_{\rm v}^2 n_{\rm o} N_{\rm aux})$ \\
  & $\widetilde F_{jb}
  = F_{jb} + \sum_Q J^Q_{bj}X^Q
  - \sum_{Qi} X^Q_{ji}J^Q_{bi}$ &
  $\mathcal{O}(n_{\rm o}^2 n_{\rm v} N_{\rm aux})$ \\

  \vspace{-2mm}\\
  4. & Form effective Fock blocks $\widehat F$ &
  $\mathcal{O}(n_{\rm o}n_{\rm v}(n_{\rm o}+n_{\rm v}))$ \\
  \hline
  & $\widehat F_{bc}
  = \widetilde F_{bc} - \sum_k \widetilde F_{kc}t_k^b$ &
  $\mathcal{O}(n_{\rm o}n_{\rm v}^2)$ \\
  & $\widehat F_{ki}
  = \widetilde F_{ki} + \sum_c \widetilde F_{kc}t_i^c$ &
  $\mathcal{O}(n_{\rm o}^2n_{\rm v})$ \\

  \vspace{-2mm}\\
  5. & Build Laplace factors $Y^{Qx}_{ai}$ &
  $\mathcal{O}(N_{\rm aux} n_{\rm o}n_{\rm v}(n_{\rm o}+n_{\rm v}))$ \\
  \hline
  & $Y^{Qx}_{ai}
  = \sqrt{w_x}
  \left[
  e^{-x\widehat F_{\rm vv}}
  \widetilde J^Q
  e^{+x\widehat F_{\rm oo}}
  \right]_{ai}$\\

  \vspace{-2mm}\\
  6. & Evaluate $\Omega$ with factorized $T_2$ &
  $\mathcal{O}(n_{\rm L} N_{\rm aux} n_{\rm o}^2 n_{\rm v})$ \\
  \hline
  & $\Omega_{ai} \mathrel{-}=
  2\sum_{Qx}
  \left(
  \sum_{jb}Y^{Qx}_{bj}\widetilde F_{jb}
  \right)Y^{Qx}_{ai}$ &
  $\mathcal{O}(n_{\rm L} N_{\rm aux} n_{\rm o}n_{\rm v})$ \\
  & $\Omega_{ai} \mathrel{+}=
  \sum_{Qxj}
  \left(
  \sum_b Y^{Qx}_{bi}\widetilde F_{jb}
  \right)Y^{Qx}_{aj}$ &
  $\mathcal{O}(n_{\rm L} N_{\rm aux} n_{\rm o}^2n_{\rm v})$ \\

  \vspace{-2mm}\\
  7. & Form active-block error $\Delta t_{IJ}^{AB}$ &
  $\mathcal{O}(n_{\rm L} N_{\rm aux} n_I^2 n_A^2)$ \\
  \hline
  & $\overline t_{IJ}^{AB}
  = -\sum_{Qx}Y^{Qx}_{AI}Y^{Qx}_{BJ}$ & \\
  & $\Delta t_{IJ}^{AB}
  = t_{IJ}^{AB,\rm HL} - \overline t_{IJ}^{AB}$ & \\

  \vspace{-2mm}\\
  8. & Build projected active-correction sources &
  $\mathcal{O}(N_{\rm frag}^4 N)$ \\
  \hline
  & $s_{iJ}^{AB}
  = -\sum_K \widehat F_{Ki}\Delta t_{KJ}^{AB}$ &
  $\mathcal{O}(n_i n_I^2 n_A^2)$ \\
  & $s_{IJ}^{aB}
  = \sum_C \widehat F_{aC}\Delta t_{JI}^{BC}$ &
  $\mathcal{O}(n_I^2 n_a n_A^2)$ \\

  \vspace{-2mm}\\
  9. & Solve projected Eq.~$\mathrm{T2\_error}$ for retained classes only &
  $\mathcal{O}(n_{\rm it} N_{\rm frag}^4 N)$ \\
  \hline
  & Unknowns:
  $\Delta t_{iJ}^{AB}$ and $\Delta t_{IJ}^{aB}$ & \\
  & Solve
  $\mathcal{R}_{iJ}^{AB}=0$ and
  $\mathcal{R}_{IJ}^{aB}=0$
  with GMRES/preconditioning & \\

  \vspace{-2mm}\\
  10. & Update $\Omega$ from retained active corrections &
  $\mathcal{O}(N_{\rm frag}^3 N)$ \\
  \hline
  & $\Omega_{Ai} \mathrel{+}=
  \sum_{JB}
  \left(2\Delta t_{iJ}^{AB}
  -\Delta t_{iJ}^{BA}\right)\widetilde F_{JB}$ & \\
  & $\Omega_{aI} \mathrel{+}=
  \sum_{JB}
  \left(2\Delta t_{IJ}^{aB}
  -\Delta t_{JI}^{aB}\right)\widetilde F_{JB}$ & \\

  \vspace{-2mm}\\
  11. & Update $t_1$ and keep $T_2$ in factorized/projected form &
  $\mathcal{O}(n_{\rm o}n_{\rm v})$ \\
  \hline
  & $t_i^a \leftarrow t_i^a - \Omega_{ai}/\epsilon_i^a$ & \\
  & Store $Y^{Qx}_{ai}$, $\Delta t_{iJ}^{AB}$, and $\Delta t_{IJ}^{aB}$ only & \\

  \bottomrule
  \end{tabular}
  \caption{\label{tabl:LLworking equations}Working equations and scaling for the factorized Laplace low-level method with projected active $T_2$ correction.}
  \end{table}

  For Step 9, I would define the residuals below the table:

  \begin{align}
  \mathcal{R}_{iJ}^{AB}
  &=
  \sum_C \widehat F_{BC}\Delta t_{iJ}^{AC}
  +\sum_C \widehat F_{AC}\Delta t_{iJ}^{CB}
  -\sum_k \widehat F_{ki}\Delta t_{kJ}^{AB}
  -\sum_K \widehat F_{KJ}\Delta t_{iK}^{AB}
  +s_{iJ}^{AB},
  \\
  \mathcal{R}_{IJ}^{aB}
  &=
  \sum_C \widehat F_{BC}\Delta t_{IJ}^{aC}
  +\sum_c \widehat F_{ac}\Delta t_{IJ}^{cB}
  -\sum_K \widehat F_{KI}\Delta t_{KJ}^{aB}
  -\sum_K \widehat F_{KJ}\Delta t_{IK}^{aB}
  +s_{IJ}^{aB}.
  \end{align}

\section{HT$_2^2$ Goldstone diagrams}

In this section, we provide the Goldstone diagrams for the HT$_2^2$ contractions. These diagrams were analyzed in the context of screened interaction construction in the main text.

\begin{table*}[t]
\centering
\caption{Goldstone diagrams appearing from the HT$_2^2$ contraction were used in the analysis of the screened interaction. Diagrams A and B are factorizable when DF is employed, while diagrams C, D, E, and F are not.}
\label{tab:rccd-diagrams}
\setlength{\tabcolsep}{8pt}
\renewcommand{\arraystretch}{1.15}
\begin{tabular}{@{}c@{\hspace{0.5em}}c@{\hspace{1.5em}}c@{\hspace{0.5em}}c@{}}
\textbf{A} &
\begin{tikzpicture}[x=1.15cm,y=1.15cm]

  \coordinate (Lext) at (-2.15,0);
  \coordinate (Lint) at (-0.65,0);
  \coordinate (Rint) at ( 0.65,0);
  \coordinate (Rext) at ( 2.15,0);
  \coordinate (VL) at (-0.65,1.55);
  \coordinate (VR) at ( 0.65,1.55);

  \draw[cluster]
    (Lext) -- (Lint)
    node[midway,below=5pt,oplab] {$T_2$};

  \draw[cluster]
    (Rint) -- (Rext)
    node[midway,below=5pt,oplab] {$T_2$};

  \draw[interaction]
    (VL) -- (VR)
    node[midway,above=5pt,oplab] {$\tilde v$};

  \draw[fermion]
    (Lext)
    .. controls (-2.55,0.45) and (-2.55,1.10)
    .. (-2.45,1.55);

  \draw[fermion]
    (-1.85,1.55)
    .. controls (-1.75,1.10) and (-1.75,0.45)
    .. (Lext);

  \node[lab,above] at (-2.45,1.55) {$a$};
  \node[lab,above] at (-1.85,1.55) {$i$};

  \draw[fermion]
    (Lint)
    .. controls (-1.05,0.45) and (-1.05,1.10)
    .. (VL);

  \draw[fermion]
    (VL)
    .. controls (-0.25,1.10) and (-0.25,0.45)
    .. (Lint);

  \node[lab,left=1pt] at (-1.03,0.88) {$c$};
  \node[lab]          at (-0.50,0.58) {$k$};

  \draw[fermion]
    (Rint)
    .. controls (0.25,0.45) and (0.25,1.10)
    .. (VR);

  \draw[fermion]
    (VR)
    .. controls (1.05,1.10) and (1.05,0.45)
    .. (Rint);

  \node[lab]           at (0.50,1.02) {$d$};
  \node[lab,right=1pt] at (1.03,0.88) {$l$};

  \draw[fermion]
    (Rext)
    .. controls (1.75,0.45) and (1.75,1.10)
    .. (1.85,1.55);

  \draw[fermion]
    (2.45,1.55)
    .. controls (2.55,1.10) and (2.55,0.45)
    .. (Rext);

  \node[lab,above] at (1.85,1.55) {$b$};
  \node[lab,above] at (2.45,1.55) {$j$};

  \foreach \p in {Lext,Lint,Rint,Rext,VL,VR}
    \node[vertex] at (\p) {};

\end{tikzpicture}
& \textbf{B} &
\begin{tikzpicture}[x=1.15cm,y=1.15cm]

  \coordinate (Lext) at (-2.15,0);
  \coordinate (Lint) at (-0.65,0);
  \coordinate (Rint) at ( 0.65,0);
  \coordinate (Rext) at ( 2.15,0);
  \coordinate (VL) at (-0.65,1.55);
  \coordinate (VR) at ( 0.65,1.55);

  \draw[cluster]
    (Lext) -- (Lint)
    node[midway,below=5pt,oplab] {$T_2$};

  \draw[cluster]
    (Rint) -- (Rext)
    node[midway,below=5pt,oplab] {$T_2$};

  \draw[interaction]
    (VL) -- (VR)
    node[midway,above=5pt,oplab] {$\tilde v$};

  \draw[fermion]
    (Lext)
    .. controls (-2.55,0.45) and (-2.55,1.10)
    .. (-2.45,1.55);

  \node[lab,above] at (-2.45,1.55) {$a$};

  \draw[fermion]
    (VL) -- (Lext);

  \draw[fermion]
    (-1.05,1.55)
    .. controls (-1.05,1.10) and (-1.05,0.45)
    .. (Lint);

  \draw[fermion]
    (Lint)
    .. controls (-.25,0.45) and (-.25,1.10)
    .. (VL);

  \node[lab,left=1pt] at (-1.3,0.88) {$k$};
  \node[lab,above] at (-1.05,1.55) {$i$};
  \node[lab] at (-0.50,0.58) {$c$};

  \draw[fermion]
    (Rint)
    .. controls (0.25,0.45) and (0.25,1.10)
    .. (VR);

  \draw[fermion]
     (VR) -- (Rext);

  \draw[fermion]
    (1.05,1.55)
    .. controls (1.05,1.10) and (1.05,0.45)
    .. (Rint);

  \node[lab] at (0.5,1.02) {$d$};
  \node[lab,right=1pt] at (1.5,0.88) {$l$};
  \node[lab,above] at (1.05,1.55) {$j$};

  \draw[fermion]
    (Rext)
    .. controls (2.55,.45) and (2.55,1.10)
    .. (2.45,1.55);

  \node[lab,above] at (2.45,1.55) {$b$};

  \foreach \p in {Lext,Lint,Rint,Rext,VL,VR}
    \node[vertex] at (\p) {};

\end{tikzpicture}
\\[1.0em]
\textbf{C} &
\begin{tikzpicture}[x=1.15cm,y=1.15cm]

  \coordinate (Lext) at (-2.15,0);
  \coordinate (Lint) at (-0.65,0);
  \coordinate (Rint) at ( 0.65,0);
  \coordinate (Rext) at ( 2.15,0);
  \coordinate (VL) at (-0.65,1.55);
  \coordinate (VR) at ( 0.65,1.55);

  \draw[cluster]
    (Lext) -- (Lint)
    node[midway,below=5pt,oplab] {$T_2$};

  \draw[cluster]
    (Rint) -- (Rext)
    node[midway,below=5pt,oplab] {$T_2$};

  \draw[interaction]
    (VL) -- (VR)
    node[midway,above=5pt,oplab] {$\tilde v$};

  \draw[fermion]
    (VL) -- (Lext);

  \draw[line width=0.9pt,
    postaction={decorate},
    decoration={
      markings,
      mark=at position 0.35 with
        {\arrow{Stealth[length=2.2mm,width=1.5mm]}}
    }]
    (Lint) -- (VR);

  \draw[line width=0.9pt,
    postaction={decorate},
    decoration={
      markings,
      mark=at position 0.72 with
        {\arrow{Stealth[length=2.2mm,width=1.5mm]}}
    }]
    (VL) -- (Rint);

  \draw[fermion]
    (VR) -- (Rext);

  \draw[fermion]
    (Lext)
    .. controls (-2.55,0.45) and (-2.55,1.10)
    .. (-2.45,1.55);

  \draw[fermion]
    (-1.05,1.55)
    .. controls (-1.05,1.10) and (-1.05,0.45)
    .. (Lint);

  \draw[fermion]
    (1.05,1.55)
    .. controls (1.05,1.10) and (1.05,0.45)
    .. (Rint);

  \draw[fermion]
    (Rext)
    .. controls (2.55,0.45) and (2.55,1.10)
    .. (2.45,1.55);

  \node[lab,above] at (-2.45,1.55) {$a$};
  \node[lab,above] at (-1.05,1.55) {$i$};
  \node[lab,above] at (1.05,1.55) {$j$};
  \node[lab,above] at (2.45,1.55) {$b$};

  \node[lab,left=1pt] at (-1.30,0.88) {$k$};
  \node[lab] at (-0.50,0.58) {$c$};
  \node[lab] at (0.50,1.02) {$d$};
  \node[lab,right=1pt] at (1.50,0.88) {$l$};

  \foreach \p in {Lext,Lint,Rint,Rext,VL,VR}
    \node[vertex] at (\p) {};

\end{tikzpicture}
& \textbf{D} &
\begin{tikzpicture}[x=1.15cm,y=1.15cm]

  \coordinate (Lext) at (-2.15,0);
  \coordinate (Lint) at (-0.65,0);
  \coordinate (Rint) at ( 0.65,0);
  \coordinate (Rext) at ( 2.15,0);
  \coordinate (VL) at (-0.65,1.55);
  \coordinate (VR) at ( 0.65,1.55);

  \draw[cluster]
    (Lext) -- (Lint)
    node[midway,below=5pt,oplab] {$T_2$};

  \draw[cluster]
    (Rint) -- (Rext)
    node[midway,below=5pt,oplab] {$T_2$};

  \draw[interaction]
    (VL) -- (VR)
    node[midway,above=5pt,oplab] {$\tilde v$};

  \draw[fermion]
    (Lext) -- (VL);

  \draw[line width=0.9pt,
    postaction={decorate},
    decoration={
      markings,
      mark=at position 0.35 with
        {\arrow{Stealth[length=2.2mm,width=1.5mm]}}
    }]
    (Lint) -- (VR);

  \draw[line width=0.9pt,
    postaction={decorate},
    decoration={
      markings,
      mark=at position 0.72 with
        {\arrow{Stealth[length=2.2mm,width=1.5mm]}}
    }]
    (VL) -- (Rint);

  \draw[fermion]
    (VR) -- (Rext);

  \draw[fermion]
    (-2.45,1.55)
    .. controls (-2.55,1.10) and (-2.55,0.45)
    .. (Lext);

  \draw[fermion]
    (-1.05,1.55)
    .. controls (-1.05,1.10) and (-1.05,0.45)
    .. (Lint);

  \draw[fermion]
    (Rint)
    .. controls (1.05,0.45) and (1.05,1.10)
    .. (1.05,1.55);

  \draw[fermion]
    (Rext)
    .. controls (2.55,0.45) and (2.55,1.10)
    .. (2.45,1.55);

  \node[lab,above] at (-2.45,1.55) {$i$};
  \node[lab,above] at (-1.05,1.55) {$j$};
  \node[lab,above] at (1.05,1.55) {$a$};
  \node[lab,above] at (2.45,1.55) {$b$};

  \node[lab,left=1pt] at (-1.30,0.88) {$k$};
  \node[lab] at (-0.50,0.58) {$c$};
  \node[lab] at (0.50,1.02) {$d$};
  \node[lab,right=1pt] at (1.50,0.88) {$l$};

  \foreach \p in {Lext,Lint,Rint,Rext,VL,VR}
    \node[vertex] at (\p) {};

\end{tikzpicture}
\\[1.0em]
\textbf{E} &
\begin{tikzpicture}[x=1.15cm,y=1.15cm]

  \coordinate (Lext) at (-2.15,0);
  \coordinate (Lint) at (-0.65,0);
  \coordinate (Rint) at ( 0.65,0);
  \coordinate (Rext) at ( 2.15,0);
  \coordinate (VL) at (-0.65,1.55);
  \coordinate (VR) at ( 0.65,1.55);

  \draw[cluster]
    (Lext) -- (Lint)
    node[midway,below=5pt,oplab] {$T_2$};

  \draw[cluster]
    (Rint) -- (Rext)
    node[midway,below=5pt,oplab] {$T_2$};

  \draw[interaction]
    (VL) -- (VR)
    node[midway,above=5pt,oplab] {$\tilde v$};

  \draw[fermion]
    (VL) -- (Lint);

  \draw[line width=0.9pt,
    postaction={decorate},
    decoration={
      markings,
      mark=at position 0.35 with
        {\arrow{Stealth[length=2.2mm,width=1.5mm]}}
    }]
    (Lint) -- (VR);

  \draw[line width=0.9pt,
    postaction={decorate},
    decoration={
      markings,
      mark=at position 0.72 with
        {\arrow{Stealth[length=2.2mm,width=1.5mm]}}
    }]
    (Rint) -- (VL);

  \draw[fermion]
    (VR) -- (Rext);

  \draw[fermion]
    (Lext)
    .. controls (-2.55,0.45) and (-2.55,1.10)
    .. (-2.45,1.55);

  \draw[fermion]
    (-1.85,1.55)
    .. controls (-1.75,1.10) and (-1.75,0.45)
    .. (Lext);

  \draw[fermion]
    (1.05,1.55)
    .. controls (1.05,1.10) and (1.05,0.45)
    .. (Rint);

  \draw[fermion]
    (Rext)
    .. controls (2.55,0.45) and (2.55,1.10)
    .. (2.45,1.55);

  \node[lab,above] at (-2.45,1.55) {$a$};
  \node[lab,above] at ( -1.85,1.55) {$i$};
  \node[lab,above] at ( 1.05,1.55) {$j$};
  \node[lab,above] at ( 2.45,1.55) {$b$};

  \node[lab,left=1pt] at (-0.65,0.78) {$k$};
  \node[lab]          at ( -0.05,0.42) {$c$};
  \node[lab]          at (-0.05,1.20) {$d$};
  \node[lab,right=1pt] at (1.50,0.88) {$l$};

  \foreach \p in {Lext,Lint,Rint,Rext,VL,VR}
    \node[vertex] at (\p) {};

\end{tikzpicture}
& \textbf{F} &
\begin{tikzpicture}[x=1.15cm,y=1.15cm]

  \coordinate (Lext) at (-2.15,0);
  \coordinate (Lint) at (-0.65,0);
  \coordinate (Rint) at ( 0.65,0);
  \coordinate (Rext) at ( 2.15,0);
  \coordinate (VL) at (-0.65,1.55);
  \coordinate (VR) at ( 0.65,1.55);

  \draw[cluster]
    (Lext) -- (Lint)
    node[midway,below=5pt,oplab] {$T_2$};

  \draw[cluster]
    (Rint) -- (Rext)
    node[midway,below=5pt,oplab] {$T_2$};

  \draw[interaction]
    (VL) -- (VR)
    node[midway,above=5pt,oplab] {$\tilde v$};

  \draw[fermion]
    (VL) -- (Lint);

  \draw[line width=0.9pt,
    postaction={decorate},
    decoration={
      markings,
      mark=at position 0.35 with
        {\arrow{Stealth[length=2.2mm,width=1.5mm]}}
    }]
    (Lint) -- (VR);

  \draw[line width=0.9pt,
    postaction={decorate},
    decoration={
      markings,
      mark=at position 0.72 with
        {\arrow{Stealth[length=2.2mm,width=1.5mm]}}
    }]
    (Rint) -- (VL);

  \draw[fermion]
    (VR) -- (Rint);

  \draw[fermion]
    (Lext)
    .. controls (-2.55,0.45) and (-2.55,1.10)
    .. (-2.45,1.55);

  \draw[fermion]
    (-1.85,1.55)
    .. controls (-1.75,1.10) and (-1.75,0.45)
    .. (Lext);

  \draw[fermion]
    (Rext)
    .. controls (2.55,0.45) and (2.55,1.10)
    .. (2.45,1.55);

  \draw[fermion]
    (1.85,1.55)
    .. controls (1.75,1.10) and (1.75,0.45)
    .. (Rext);

  \node[lab,above] at (-2.45,1.55) {$a$};
  \node[lab,above] at ( -1.85,1.55) {$i$};
  \node[lab,above] at ( 1.85,1.55) {$j$};
  \node[lab,above] at ( 2.45,1.55) {$b$};

  \node[lab,left=1pt] at (-0.65,0.78) {$k$};
  \node[lab]          at ( -0.05,0.42) {$c$};
  \node[lab]          at (-0.05,1.20) {$d$};
  \node[lab,right=1pt] at (0.65,0.88) {$l$};

  \foreach \p in {Lext,Lint,Rint,Rext,VL,VR}
    \node[vertex] at (\p) {};

\end{tikzpicture}
\end{tabular}
\end{table*}

\newpage

\twocolumngrid

\bibliographystyle{apsrev4-2}
\bibliography{lib, misc}

\end{document}